\documentclass[
amsmath,
amssymb,
aps,
prstab,
floatfix,
twocolumn,
]{revtex4}

\usepackage{subfigure}
\usepackage{graphicx}
\usepackage{dcolumn}
\usepackage{bm}
\usepackage{multirow}

\usepackage[usenames]{color}
\usepackage{hyperref}

\begin{document}


\title{Transverse phase space characterisation in the CLARA FE accelerator test facility at Daresbury Laboratory}

\author{A.\,Wolski}
\email{a.wolski@liverpool.ac.uk}
\affiliation{University of Liverpool, Liverpool, UK, and the Cockcroft Institute, Daresbury, UK.}

\author{D.C.\,Christie}
\affiliation{University of Liverpool, Liverpool, UK, and the Cockcroft Institute, Daresbury, UK.}

\author{B.L.\,Militsyn}
\affiliation{STFC/ASTeC, Daresbury Laboratory, Daresbury, UK.}

\author{D.J.\,Scott}
\affiliation{STFC/ASTeC, Daresbury Laboratory, Daresbury, UK.}

\author{H.\,Kockelbergh}
\affiliation{STFC/ASTeC, Daresbury Laboratory, Daresbury, UK.}

\date{\today}

\begin{abstract}
We compare three techniques for characterising the transverse phase space distribution of
the beam in CLARA FE (the Compact Linear Accelerator for Research and Applications Front End, at
Daresbury Laboratory, UK): emittance and optics measurements using screens at three separate
beamline locations; quadrupole scans; and phase space tomography.  We find that where the
beam distribution has significant structure (as in the case of CLARA FE at the time the measurements
presented here were made) tomography analysis is the most reliable way to obtain a
meaningful characterisation of the transverse beam properties.  We present the first experimental
results from four-dimensional phase space tomography: our results show that this technique can
provide an insight into beam properties that are of importance for optimising machine performance.
\end{abstract}

\maketitle



\section{\label{sec:intro}Introduction}



Knowledge of transverse beam emittance and optical properties are essential for the commissioning and
performance optimisation of many accelerator facilities.  There are well-established techniques for emittance
and optics measurements, often based on observation of changes in beam size in response to changes in strength
of focusing (quadrupole) magnets, or observation of the beam size at different locations along a beam line
\cite{wiedemann2007,minty2003}.  Beam phase space tomography is also an established method for providing
detailed information about the phase space distribution
\cite{mckee1995,yakimenko2003,stratakis2003,stratakis2007,xiang2009,xing2018,ji2019,rohrs2009}.
In this paper, we report the results
of studies on CLARA FE (Compact Linear Accelerator for Research and Applications, Front End) at Daresbury Laboratory
\cite{claracdr,clara1}, aimed at characterising the transverse emittance and
optical properties of the electron beam.  The results of three different measurement techniques are compared,
namely: beam-size measurements at three different locations along the beamline (``three-screen analysis'');
measurement in the change of the beam size in response to the change in quadrupole strengths (``quadrupole scan'');
and beam phase space tomography, with which we demonstrate for the first time reconstruction of the four-dimensional
transverse phase space.
At the time that the studies were carried out, the beam in CLARA FE had significant detailed structure in the
phase space distribution (i.e.~the phase space distribution could not be described by a simple Gaussian).
We find that in these circumstances, phase space tomography provides the most reliable characterisation
of the transverse beam properties.  Quadrupole scans can provide some useful information, but the
results from three-screen analysis can be unreliable.  Our studies of phase space tomography include
the first experimental demonstration of beam tomography in four-dimensional phase space \cite{friedman2003, hock2013}.
We find that this technique can provide an insight into coupling in the beam, which can be of value for
optimising machine performance \cite{zheng2019}.

This paper is organised as follows.  In Section~\ref{sec:emittancecalculation} we briefly review the definitions
that we use for the emittances and optics functions in coupled beams, and the methods that we use for calculating
these quantities.  In Section~\ref{sec:experimentalprocedures} we describe the measurement procedures in
CLARA FE.  The three-screen analysis method is discussed in Section~\ref{sec:threescreen}, where simulation and
experimental results for a single measurement case are presented.  The results show some limitations of the
technique, and we discuss in particular why it does not produce reliable results when
the beam has a complicated structure in phase space (i.e.~the distribution is not a simple Gaussian).
In Section~\ref{sec:quadscan} we describe the quadrupole scan analysis method, including application to measurement
of the full covariance matrix in two (transverse) degrees of freedom.  The quadrupole scan technique has some
advantages over the three-screen analysis, but neither method can determine the detailed structure of the beam
distribution in phase space. Such information can be provided by the final analysis technique, phase space
tomography, which is considered in Section~\ref{sec:tomography}.  We describe the phase space tomography
technique, including the use of normalised phase space \cite{hock2011}, and show how tomography can be
applied to determine the beam distribution in four-dimensional phase space \cite{hock2013}.  Simulation results are presented
to validate the technique, and some experimental results are again presented.  In Section~\ref{sec:applications}
we show the application of phase space tomography to provide a detailed characterisation of the beam
in CLARA FE under a range of machine conditions, looking at the dependence of emittance and optics
(including coupling) on strength of the electron source solenoid, bucking coil, and bunch charge.  Given the detailed
structure generally present in the phase space distribution of the beam, phase space tomography provides
important insights into the beam properties and behaviour that would not be obtained from the three-screen
or quadrupole scan analysis techniques.  Tomography in four-dimensional phase space provides,
in particular, information on beam coupling that is of value for optimising machine performance.
Finally, in Section~\ref{sec:conclusions}, we
summarise the key results, discuss the main conclusions, and consider appropriate directions for further work.

\section{Normal mode emittances and optical functions\label{sec:emittancecalculation}}
Since various definitions of beam emittance are used in different contexts, we briefly review
the definition we use for the studies presented here, considering in particular the case
where there is coupling in the beam.
For clarity, however, we begin with the case of a single degree of freedom.
Considering, for example, the transverse horizontal direction, the covariance matrix at a specified point in a beamline can be written:
\begin{equation}
\Sigma = \left( \begin{array}{cc}
\langle x^2 \rangle & \langle x p_x \rangle \\
\langle x p_x \rangle & \langle p_x^2 \rangle \\
\end{array} \right),
\label{sigmamatrix1dof}
\end{equation}
where $x$ represents the transverse horizontal co-ordinate of a single particle at the
specified location, $p_x$ is the
horizontal momentum $P_x$ (at the same location) divided by a chosen reference momentum $P_0$, and the brackets $\langle \, \rangle$
indicate an average over all particles in the beam.  Note that we assume there is no dispersion in
the beamline, so that the trajectory of the beam (nominally passing through the centre of each quadrupole)
is independent of its energy: for the present studies in CLARA FE, since the layout from the electron source to the end of the section where the emittance measurements are performed
is a straight line, this will be a good approximation.

The horizontal (geometric) emittance $\epsilon_x$ and
optical functions (Courant--Snyder parameters $\beta_x$ and $\alpha_x$ and $\gamma_x$) can be calculated from:
\begin{eqnarray}
\epsilon_x & = & \sqrt{\langle x^2 \rangle \langle p_x^2 \rangle - \langle x p_x \rangle^2},
\label{eqnemittance} \\
\beta_x & = & \frac{\langle x^2 \rangle}{\epsilon_x},
\label{eqnbetafunction} \\
\alpha_x & = & - \frac{\langle x p_x \rangle}{\epsilon_x},
\label{eqnalphafunction} \\
\gamma_x & = & \frac{\langle p_x^2 \rangle}{\epsilon_x}.
\label{eqngammafunction}
\end{eqnarray}
These relations imply that:
\begin{equation}
\gamma_x\langle x^2 \rangle +
2\alpha_x\langle x p_x \rangle +
\beta_x\langle p_x^2 \rangle = 2\epsilon_x.
\label{emittanceellipse}
\end{equation}
Equation (\ref{emittanceellipse}) defines an ``emittance ellipse'' in phase space.
It is straightforward to extend these results to the vertical direction, to find the vertical
emittance and Courant--Snyder parameters.

In considering only a single degree of freedom, we assume that there is no transverse
coupling in the beam or in the beamline,
so that the transverse horizontal and vertical motions may be treated independently.  Coupling in the
beam will be characterised by non-zero values for cross-plane elements (such as $\langle x y \rangle$,
for example) in the $4\times 4$ covariance matrix.  Coupling in the beamline will arise from skew components
in the quadrupoles (for example, from some alignment error in the form of a tilt of the magnet around the
beam axis) or from a solenoid field either at the source or further down the beamline.
If there is coupling in the beam, then the emittance calculated using
(\ref{eqnemittance}) will not be the most useful quantity, since it will not be constant as the beam travels along
the beamline.  The conserved quantities where coupling is present are the normal mode emittances (or eigenemittances) $\epsilon_\mathrm{I}$
and $\epsilon_\mathrm{II}$, where $\pm i \epsilon_\mathrm{I,II}$ are the eigenvalues of $\Sigma S$,
with $\Sigma$ the $4\times 4$ covariance matrix, and $S$ the antisymmetric matrix:
\begin{equation}
S = \left( \begin{array}{cccc}
0 & 1 & 0 & 0 \\
-1 & 0 & 0 & 0 \\
0 & 0 & 0 & 1 \\
0 & 0 & -1 & 0
\end{array} \right).
\label{unitantisymmetric}
\end{equation}

Various formalisms have been developed for generalising the emittance and
Courant--Snyder parameters from one to two (or more) coupled degrees of freedom.  Here, we use
the method presented in \cite{wolski2006}, in which the $(i,j)$ element of a covariance matrix
$\Sigma$ is related to the normal mode emittances $\epsilon_\mathrm{I}$, $\epsilon_\mathrm{II}$
and corresponding optical functions $\beta_{ij}^\mathrm{I}$, $\beta_{ij}^\mathrm{II}$ by:
\begin{equation}
\Sigma_{ij} = \sum_{k= \mathrm{I}, \mathrm{II}} \epsilon_k \beta_{ij}^k.
\label{sigmaequalsepsilonbeta}
\end{equation}
The optical functions can be obtained from the eigenvectors of $\Sigma S$.
If $U$ is a matrix constructed from the eigenvectors (arranged in columns) of $\Sigma S$, then:
\begin{equation}
\Sigma S = U \Lambda U^{-1},
\end{equation}
where $\Lambda$ is a diagonal matrix with diagonal elements corresponding to the eigenvalues
of $\Sigma S$.  If the eigenvectors and eigenvalues are arranged so that, in two degrees of freedom:
\begin{equation}
\Lambda = \left( \begin{array}{cccc}
-i \epsilon_\mathrm{I} & 0 & 0 & 0 \\
0 & i \epsilon_\mathrm{I} & 0 & 0  \\
0 & 0 & -i \epsilon_\mathrm{II} & 0 \\
0 & 0 & 0 & i \epsilon_\mathrm{II}
\end{array} \right),
\end{equation}
then the optical functions are given by:
\begin{equation}
\beta^k = U E^k U^{-1} S,
\end{equation}
where $k = \mathrm{I}, \mathrm{II}$, and:
\begin{equation}
E^\mathrm{I} = 
\left( \begin{array}{cccc}
i & 0 & 0 & 0 \\
0 & -i & 0 & 0  \\
0 & 0 & 0 & 0 \\
0 & 0 & 0 & 0
\end{array} \right),
\quad
E^\mathrm{II} = 
\left( \begin{array}{cccc}
0 & 0 & 0 & 0 \\
0 & 0 & 0 & 0  \\
0 & 0 & i & 0 \\
0 & 0 & 0 & -i
\end{array} \right).
\end{equation}
The covariance matrix $\Sigma$ can then be expressed in terms of the normal mode
emittances and optical functions using (\ref{sigmaequalsepsilonbeta}).

When there is no coupling, the normal mode emittances and optical functions correspond to the
usual quantities defined for independent degrees of freedom.  For example, where the
transverse horizontal motion is independent of the vertical and longitudinal motion,
then:
\begin{equation}
\epsilon_\mathrm{I} = \sqrt{\langle x^2 \rangle \langle p_x^2 \rangle - \langle x p_x \rangle^2} = \epsilon_x,
\end{equation}
and:
\begin{equation}
\beta^\mathrm{I}_{1,1} = \beta_x, \quad \textrm{ and } \quad
\beta^\mathrm{I}_{1,2} = -\alpha_x.
\end{equation}

Finally, we note that if the optical functions $\beta^k_{ij}$ at a given point $s_1$ in a
beamline are known, the optical functions at any other point $s_2$ are readily computed using:
\begin{equation}
\beta^k_{ij}(s_2) = M_{21} \beta^k_{ij}(s_1) M_{21}^\mathrm{T},
\end{equation}
where $M_{21}$ is the transfer matrix from $s_1$ to $s_2$ (calculated, for example, from a
computational model of the beamline).  The normal mode emittances and optical functions
defined as described here, therefore provide convenient quantities for describing the
variation of the beam sizes $\langle x^2 \rangle$, $\langle y^2 \rangle$ (and other elements
of the covariance matrix) along a given beamline.

\section{Measurements in CLARA FE\label{sec:experimentalprocedures}}
\begin{figure*}[th]
   \centering
   \includegraphics[clip, trim=0cm 0cm 0cm 0cm, width=0.8\textwidth]{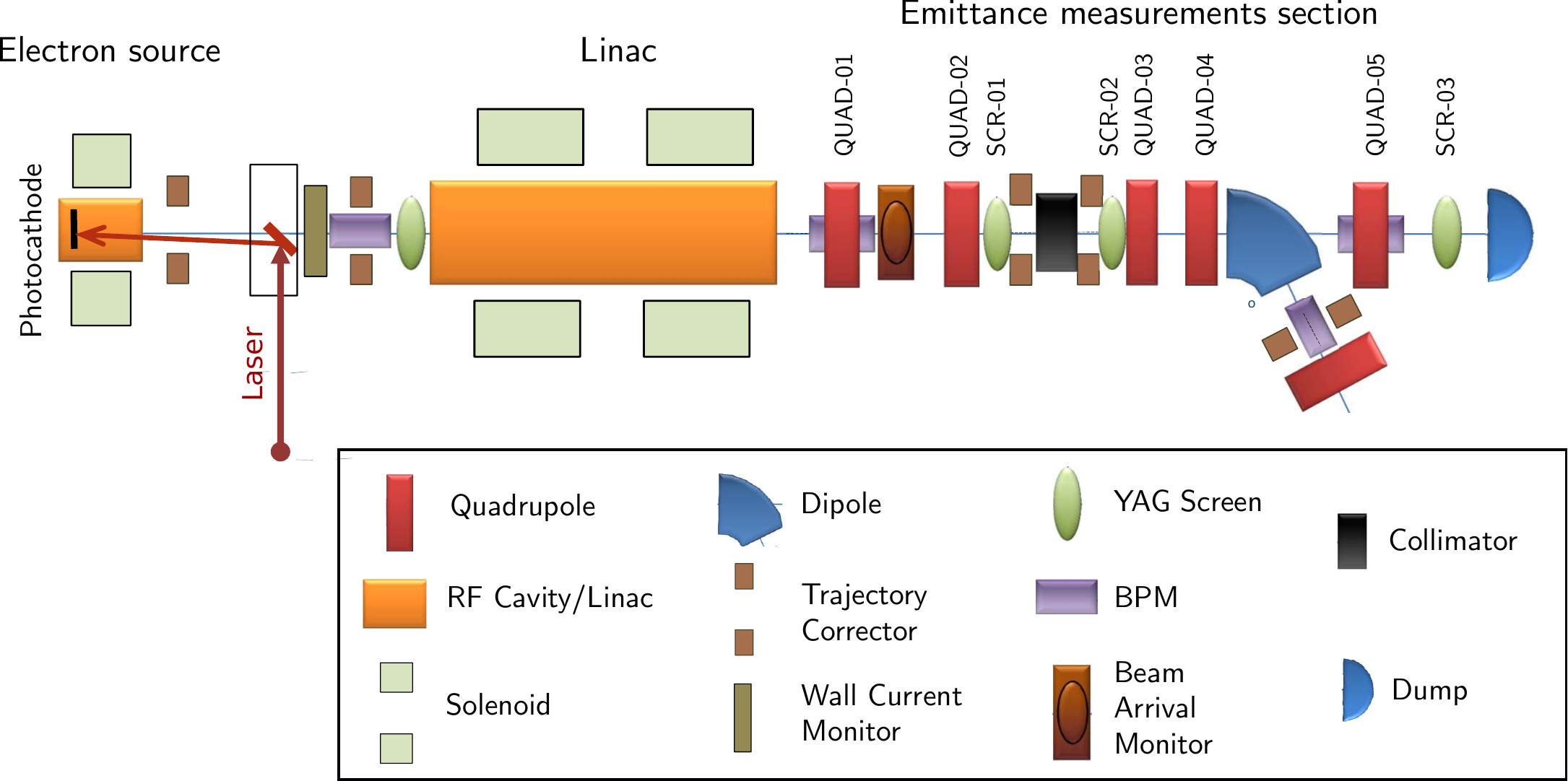}
   \caption{Layout of CLARA FE (not to scale), showing the electron source,
   magnetic elements, linac, and diagnostics.  For the emittance measurements, quadrupoles
   QUAD-01 and QUAD-02 were used to set the required optics at screen SCR-01.  For each
   measurement, beam images on SCR-03 (the Observation Point) were recorded for different
   strengths of quadupoles QUAD-03, QUAD-04 and QUAD-05. The optical functions and phase
   space distribution were computed at SCR-02 (the Reconstruction Point). Screens
   SCR-01, SCR-02 and SCR-03 are 0.61\,m, 1.53\,m and 3.20\,m from the entrance of QUAD-01,
   respectively. }
   \label{figinjectionbeamline}
\end{figure*}

Ultimately, CLARA is planned as a facility that will provide a high-quality electron beam with energy up to
250\,MeV for scientific and medical research, and for the development of new accelerator technologies
including (with the addition of an undulator section)
the testing of advanced techniques and novel modes of FEL operation.  So far, only the
front end (CLARA FE) has been constructed: this consists of a low-emittance rf photocathode electron source and a
linac reaching 35.5\,MeV/c beam momentum. The layout of CLARA FE is shown in Fig.~\ref{figinjectionbeamline}.
The electron source \cite{claragun} consists of a 2.5 cell S-band rf cavity with copper photocathode,
and can deliver short (of order a few ps) bunches at 10\,Hz repetition rate with charge in excess of 250\,pC
and with beam momentum up to 5.0\,MeV/c.  The source is driven with the third harmonic of a short
(2\,ps full-width at half maximum) pulsed Ti:Sapphire laser with a pulse energy of up to 100\,$\mu$J.
The typical size of the laser spot on the photocathode is of order 600\,$\mu$m. The source is immersed
in the field of a solenoid magnet which provides emittance compensation and focusing of the beam
in the initial section of the beamline. A bucking coil located beside the source cancels the field from the
solenoid on the photocathode in the region of the laser spot.

The studies reported in this paper are based on measurements made in the section of CLARA FE
following the linac, at a nominal beam momentum of 30\,MeV/c.  Measurements were made under a range
of conditions including various bunch charges, and different strengths of the solenoid and bucking coil at
the electron source.  Three techniques were used, to allow a comparison of the results and evaluation
of the benefits and limitations of the different methods.  The first technique, the three-screen
measurement and analysis method (described in more detail in Section~\ref{sec:threescreen}, below),
is based on observations of the transverse beam profile at three scintillating (YAG) screens, shown as
SCR-01, SCR-02 and SCR-03 in Fig.~\ref{figinjectionbeamline}.  The quadrupole scan
(Section~\ref{sec:quadscan}) and tomography (Section~\ref{sec:tomography}) methods use only
observations of the beam on SCR-03, though observations on SCR-02 were also made,
and used to validate the results.  For each of the three methods, two quadrupoles (QUAD-01 and
QUAD-02) between the end of the linac and SCR-01 were used for setting the optical functions of the
beam on SCR-01, and were kept at fixed strengths during data collection.  A collimator is located
between SCR-01 and SCR-02, but this was not used during the measurements.
For all three measurement techniques, the strengths of three quadrupoles (QUAD-03, QUAD-04
and QUAD-05 in Fig.~\ref{figinjectionbeamline}) located between SCR-02
and SCR-03 were varied.  For the three-screen analysis, only the beam sizes for one set of
magnet strengths are strictly needed to calculate the emittances and optical functions;
however, as described in Section~\ref{sec:threescreen}, measurements with different
sets of quadrupole strengths can be used to validate the results by showing the consistency for
emittance and optics values obtained for different strengths.

In the case of the quadrupole scan and tomography methods, SCR-03 provides the necessary
data, and is referred to as the ``Observation Point''.  For ease of comparison, for all three
techniques we construct the covariance matrix at SCR-02, which is referred to as the
``Reconstruction Point''.  

At each point in a quadrupole scan on CLARA FE, ten screen images were recorded on successive machine pulses
(at a rate of 10\,Hz, with a single bunch per pulse): this allows an estimate to be made of
random errors arising from pulse-to-pulse variations in beam properties.  A background image
was recorded without beam (i.e.~with the photocathode laser blocked), so that any constant
artefacts in the beam images, for example from dark current, could be subtracted.  The rms beam
sizes were calculated by projecting the image onto either the $x$ or $y$ axis, with co-ordinates
measured with respect to a centroid such that:
\begin{equation}
\langle x \rangle = \langle y \rangle = 0.
\end{equation}
Average quantities are calculated from a beam image by integration of the image intensity with an
appropriate weighting, for example:
\begin{equation}
\langle x \rangle = \frac{\iint x I(x,y) \, dx \, dy}{\iint I(x,y) \, dx \, dy},
\end{equation}
where $I(x,y)$ is the image intensity at a given point on the screen.

Between each quadrupole 
scan, the quadrupole magnets were cycled over a set range of strengths to minimise systematic
errors from hysteresis.  Remaining sources of systematic errors include calibration factors
for the magnets (when converting from coil currents to field gradients), magnet fringe fields,
calibration factors for the screens, and accelerating gradient in the linac. It was found that
better agreement between the analysis results
and direct observations (used to validate the results) could be obtained if the beam momentum in
the model used in the analysis was reduced slightly from the nominal 30\,MeV/c.  In the results
presented here, a momentum of 29.5\,MeV/c is used.  It should also be noted that some variation
in machine parameters (including rf phase and amplitude in the electron source and the
linac) is likely to have occurred during data collection, and because of the time required to
re-tune the machine it was not always possible to confirm all the parameter values between
quadrupole scans.

In principle, for each of the three measurement and analysis methods, the strengths of the quadrupoles
between SCR-02 and SCR-03 can be chosen randomly.  However, if the profile of the beam on any of the
screens becomes too large, too small, or very asymmetric (with large aspect ratio) then there can
be a large error in the calculation of the rms beam size.  Before collecting data, therefore, simulations
were performed to find sets of magnet strengths, with fixed QUAD-01 and QUAD-02 and variable
QUAD-03, QUAD-04 and QUAD-05, for which the transverse beam profiles on each of the three
screens would remain approximately circular, and with a convenient size.  It is also worth noting that,
from (\ref{eqnemittance}), a large value of $\alpha_x$ at a given location can indicate a large value
for $ \langle x p_x \rangle$ at that location: calculation of the emittance then involves taking the
difference between quantities that may be of similar magnitude, leading to a large uncertainty in
the result.  A further constraint, therefore, was to find strengths for QUAD-01 and QUAD-02 that
would provide a beam waist in $x$ and $y$ (i.e.~with $\alpha_x$ and $\alpha_y$ close to zero)
at SCR-02 (the Reconstruction Point).  Finally, quadrupole strengths were chosen to provide a
wide range of horizontal and vertical phase advances from SCR-02 to SCR-03: this is a consideration
for the tomographic analysis, and is discussed further in Section~\ref{sec:tomography}.  Simulations
to find sets of suitable strengths for all five quadrupoles were carried out in \textsc{GPT} \cite{gpt},
tracking particles from the photocathode (with nominal laser spot size and pulse length) to SCR-03, using
machine conditions matching those planned for the experiments. Space charge effects were included
\cite{gptsc1, gptsc2}, though these effects are only really significant at low momentum, upstream of the linac.


\subsection{\label{sec:threescreen}Three-screen method}



The three-screen analysis method is based on the principle that, given the rms beam
size (in either the transverse horizontal or vertical direction) at three separate
locations, and knowing the transfer matrices between those locations, it is possible
to calculate the covariance matrix characterising the phase space beam distribution.

If the transfer matrix from one location in the beamline $s_1$ to another location
$s_2$ is $M_{21}$ (with transpose $M_{21}^\mathrm{T}$), then:
\begin{equation}
\Sigma_2 = M_{21} \Sigma_1 M_{21}^\mathrm{T},
\label{eqnsigma2}
\end{equation}
where $\Sigma_1$ is the covariance matrix at $s_1$ and
$\Sigma_2$ is the covariance matrix at $s_2$.  Similarly, at a third location $s_3$:
\begin{equation}
\Sigma_3 = M_{32} \Sigma_2 M_{32}^\mathrm{T},
\label{eqnsigma3}
\end{equation}
where $\Sigma_3$ is the covariance matrix at $s_3$, and $M_{32}$ is the transfer
matrix from $s_2$ to $s_3$.
The elements of the transfer matrices can be calculated using a linear
model of the beamline, with known quadrupole strengths.

Given measured values of $\langle x_1^2 \rangle$,
$\langle x_2^2 \rangle$ and $\langle x_3^2 \rangle$ (from observation of the beam
images on the three screens), using (\ref{eqnsigma2}) and (\ref{eqnsigma3}) we
can find $\langle x_2 p_{x2} \rangle$ and $\langle p_{x2}^2 \rangle$ from:
\begin{eqnarray}
\langle x_2 p_{x2} \rangle & = &
- \frac{ b_{32}^2 \langle x_1^2 \rangle + C_+ C_- \langle x_2^2 \rangle - b_{21}^2 \langle x_3^2 \rangle}
{2 C_- b_{21} b_{32}  }, \label{eqnxpx} \\
\langle p_{x2}^2 \rangle & = &
\frac{a_{32}b_{32} \langle x_1^2 \rangle + C_- a_{21}a_{32} \langle x_2^2 \rangle - a_{21}b_{21} \langle x_3^2 \rangle}
{C_- b_{21} b_{32} }, \nonumber \\
& & \label{eqnpx2}
\end{eqnarray}
where $a_{21}$ and $b_{21}$ are (respectively) the $(1,1)$ and $(1,2)$ elements of $M_{21}^{-1}$,
$a_{32}$ and $b_{32}$ are (respectively) the $(1,1)$ and $(1,2)$ elements of $M_{32}$, and:
\begin{equation}
C_\pm = a_{32}b_{21} \pm a_{21}b_{32}.
\end{equation}

In principle, the three-screen analysis method can be extended to construct the $4\times 4$ covariance matrix
describing the beam distribution in two transverse degrees of freedom.  However, the $4\times 4$ covariance
matrix has ten independent elements, while observation of the beam profile on screens at three separate
locations provides only nine observable quantities ($\langle x^2 \rangle$,
$\langle xy \rangle$ and $\langle y^2 \rangle$ at each screen).  Therefore, using observations of the
beam profile at three screens does not provide sufficient information to determine all elements of the
$4\times 4$ covariance matrix at any point in the beamline, and it is not possible, without additional
measurements, to calculate the normal mode emittances.  The normal mode emittances can be calculated
if observations are made with different sets of quadrupole strengths between the screens (see, for example,
\cite{raimondi1993,prat2014}): we return to this point
in the discussion of the quadrupole scan technique in Section~\ref{sec:quadscan}.  Alternatively,
measurements can be made at additional longitudinal positions \cite{woodley2000}.  Measurements of the
$4\times 4$ covariance matrix (for picometer-scale emittance beams) using measurements at
multiple longitudinal locations were recently reported by Ji et al. \cite{ji2019}.

In the case of CLARA FE, we apply the three-screen analysis technique in horizontal and vertical degrees
of freedom separately (neglecting any coupling), with $s_1$, $s_2$ and $s_3$ corresponding to the locations
of screens SCR-01, SCR-02 and SCR-03 respectively (see Fig.~\ref{figinjectionbeamline}).  Results may
be validated by repeating the measurements for different strengths of the three quadrupoles between
screens SCR-02 and SCR-03: since the magnets upstream of SCR-02 remain at constant strength,
measurements for different strengths of downstream magnets should all yield the same values for the
emittances and Courant--Snyder parameters at this screen.  

\begin{figure*}[t]
\begin{center}
\includegraphics[width=0.98\columnwidth,trim={3.0cm 8.5cm 4.2cm 2.2cm},clip]{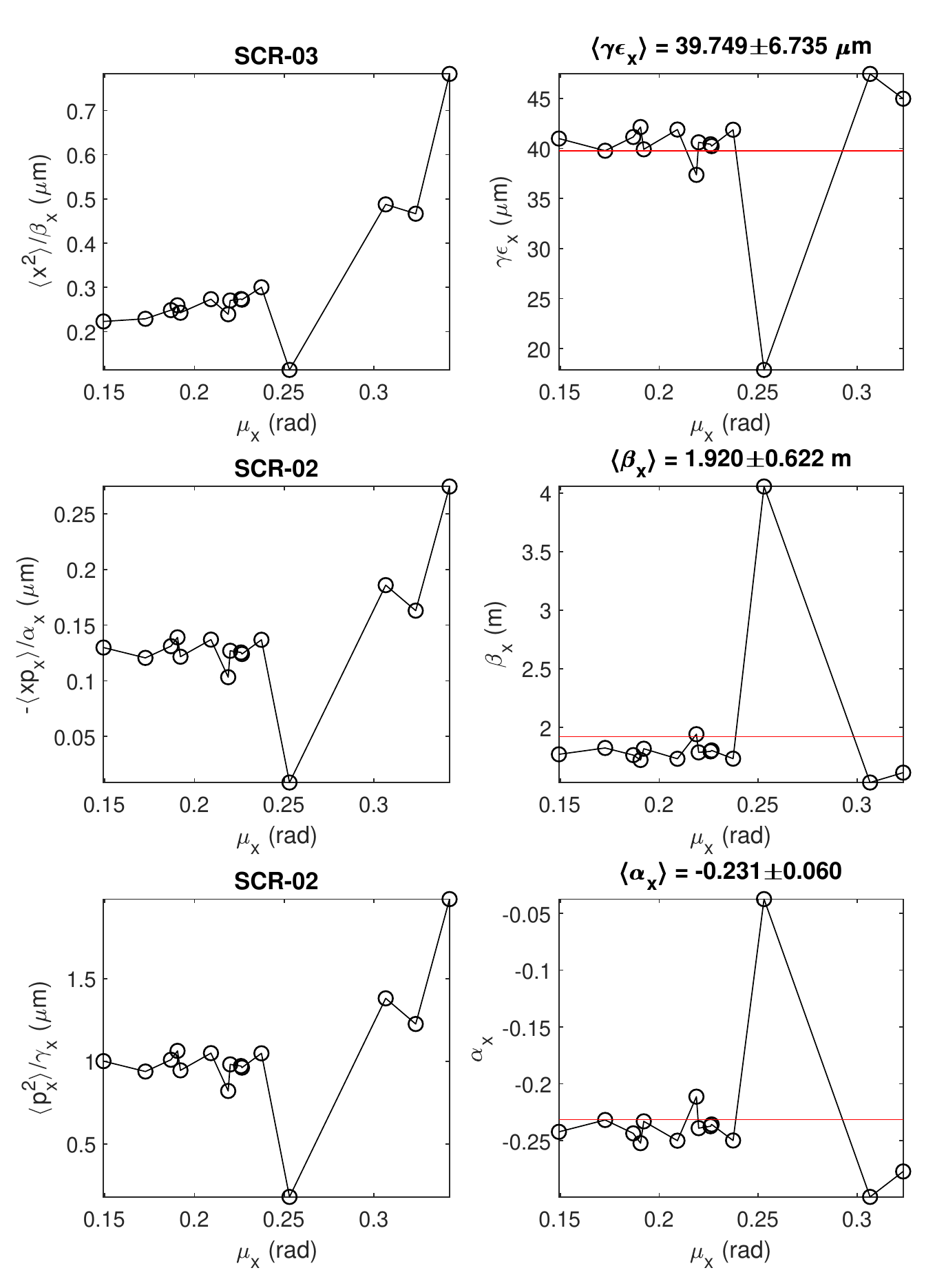}
\includegraphics[width=0.98\columnwidth,trim={3.0cm 8.5cm 4.2cm 2.2cm},clip]{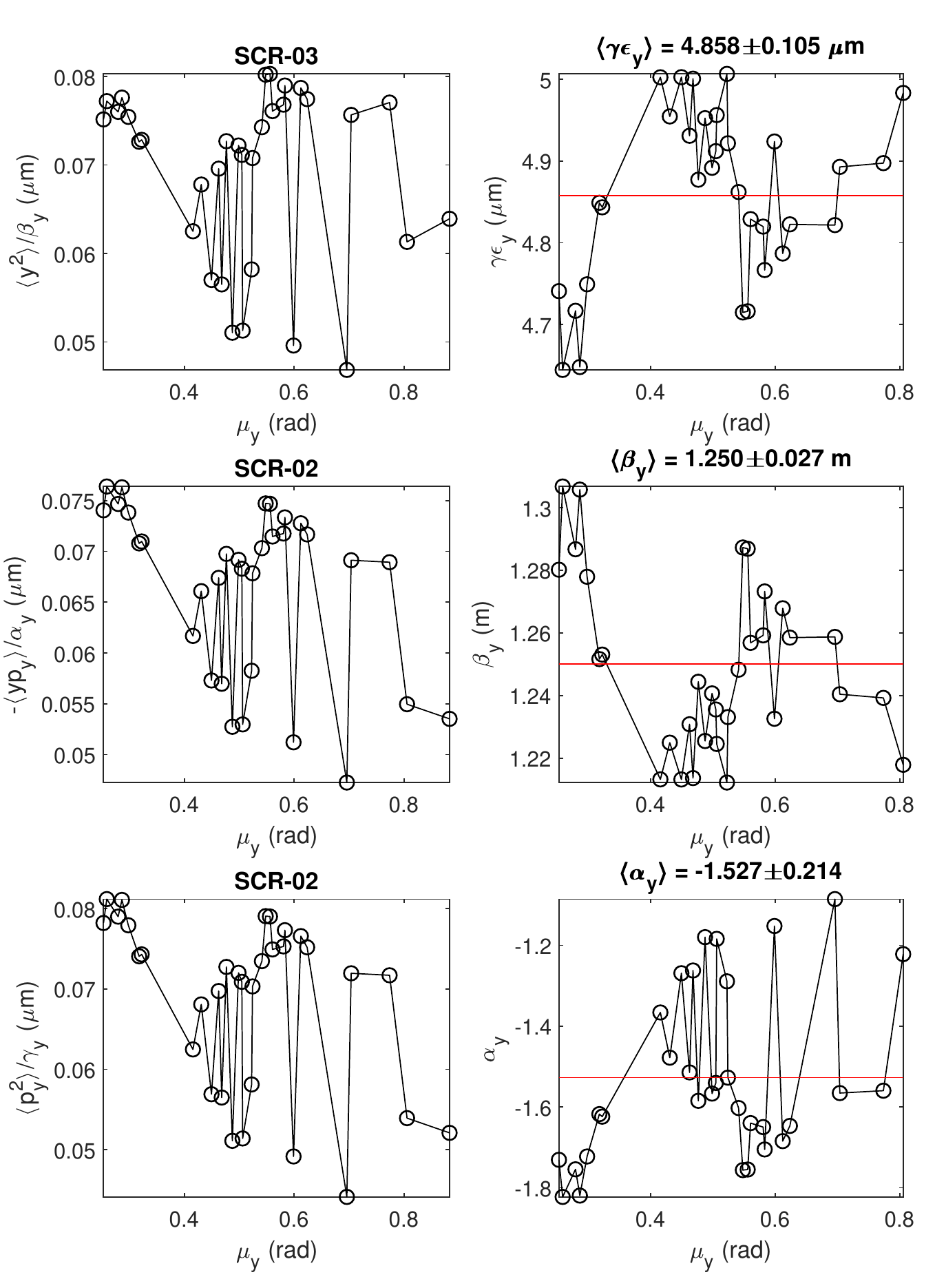}
\caption{Horizontal (left two columns) and vertical (right two columns) emittance and optics measurements
made using beam images observed at three different
locations in the beam line (screens SCR-01, SCR-02 and SCR-03).  Each point represents a different set of strengths
for the quadrupoles between SCR-02 and SCR-03. The horizontal axis shows the phase advance in the respective plane
from SCR-02 to SCR-03; the phase advance from SCR-01 and SCR-02 is fixed.  The plots in the first column show
(top) $\langle x^2 \rangle/\beta_x$ at SCR-03;
(middle) $-\langle xp_x \rangle/\alpha_x$ at SCR-02 calculated from (\ref{eqnxpx}); and
(bottom) $\langle p_x^2 \rangle/\gamma_x$ at SCR-02 calculated from (\ref{eqnpx2}).
The plots in the second column show
(top) the normalised horizontal emittance;
(middle) the horizontal beta function at SCR-02; and
(bottom) the horizontal alpha function at SCR-02.
The plots in the third and fourth columns correspond to those in the first and second columns, but for the vertical
rather than the horizontal plane.
The emittance and optical functions are calculated from the covariance matrix at SCR-02, using (\ref{eqnemittance}),
(\ref{eqnbetafunction}) and (\ref{eqnalphafunction}). Points leading to
imaginary values for the emittance are omitted.  \label{figthreescreenanalysis}}
\end{center}
\end{figure*}

Figure \ref{figthreescreenanalysis} shows a
typical set of results from the three-screen analysis, for the transverse horizontal and vertical
directions, respectively.  Elements of the covariance matrix scaled by the appropriate Courant--Snyder
parameter are plotted as a function of the phase advance from SCR-02 to SCR-03 in the respective plane
(corresponding to different strengths of the quadrupoles QUAD-03, QUAD-04 and QUAD-05).
In the case of a simple Gaussian distribution with no coupling, from
Eqs.~(\ref{eqnbetafunction})--(\ref{eqngammafunction}) we see that scaling the covariance matrix
elements by the Courant--Snyder parameters should give values that are independent of the phase
advance, and equal to the geometric emittance.  However, the results in Fig.~\ref{figthreescreenanalysis}
show significant variation in each of the scaled elements of the covariance matrix over the range of the
quadrupole scan: this is particularly evident in the horizontal direction, and is reflected in the values
calculated for the emittance and optics functions.  For the horizontal plane, a significant number of points
in the quadrupole scan lead to imaginary values for the emittance (the covariance matrix has negative
determinant), or non-physical negative values for the covariance matrix element $\langle p_x^2 \rangle$.
Simulation studies, which we discuss further below, suggest that
these features result from the non-Gaussian distribution of particles in phase space.  Theoretically, this
may be understood in terms of the way that the distribution is constructed from observations at the
three screens.  At each screen, we measure the width of a projection of the
phase space distribution onto an axis at a particular phase angle.  With three screens, we have projections
at three phase angles: if the distribution in phase space is Gaussian, this is sufficient to determine the size
and shape of the distribution (which can be described by three parameters; for example, the elements of
the covariance matrix, or the emittance and Courant--Snyder parameters).  However, a more general
distribution with a more complicated structure cannot be described by just three parameters: attempting
to do so will lead to different values for those parameters, depending on the particular phase angles chosen
for the projections.  The results of the tomography analysis presented in Section~\ref{sec:tomography}
show that at the time that the measurements reported here were made, the beam exhibited a complicated
(non-Gaussian) structure in phase space, particularly in the horizontal plane.

\begin{figure*}[th]
\begin{center}
\includegraphics[width=\columnwidth,trim={3.0cm 13.6cm 4.5cm 2.3cm},clip]{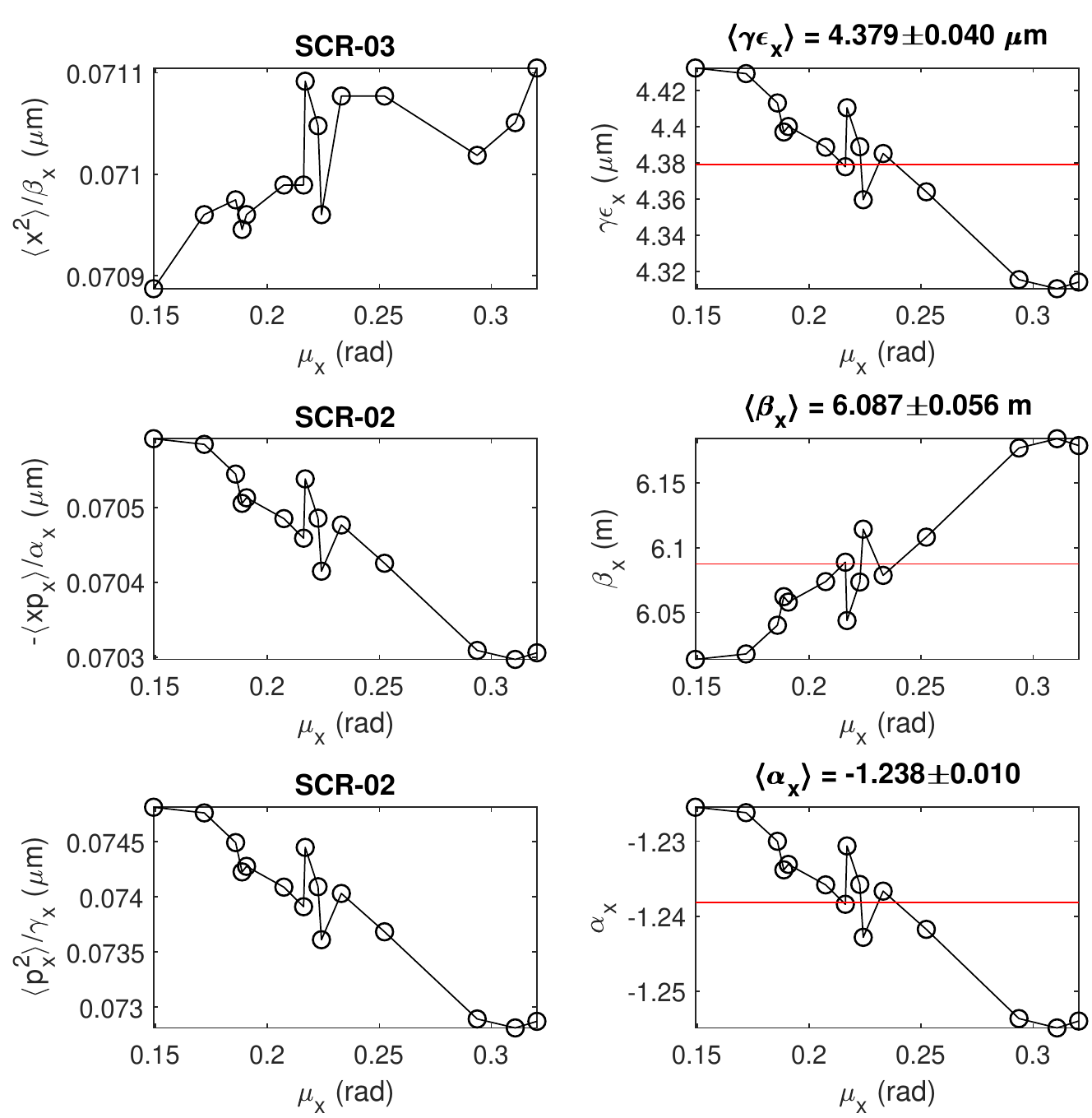}
\includegraphics[width=\columnwidth,trim={3.0cm 13.6cm 4.5cm 2.3cm},clip]{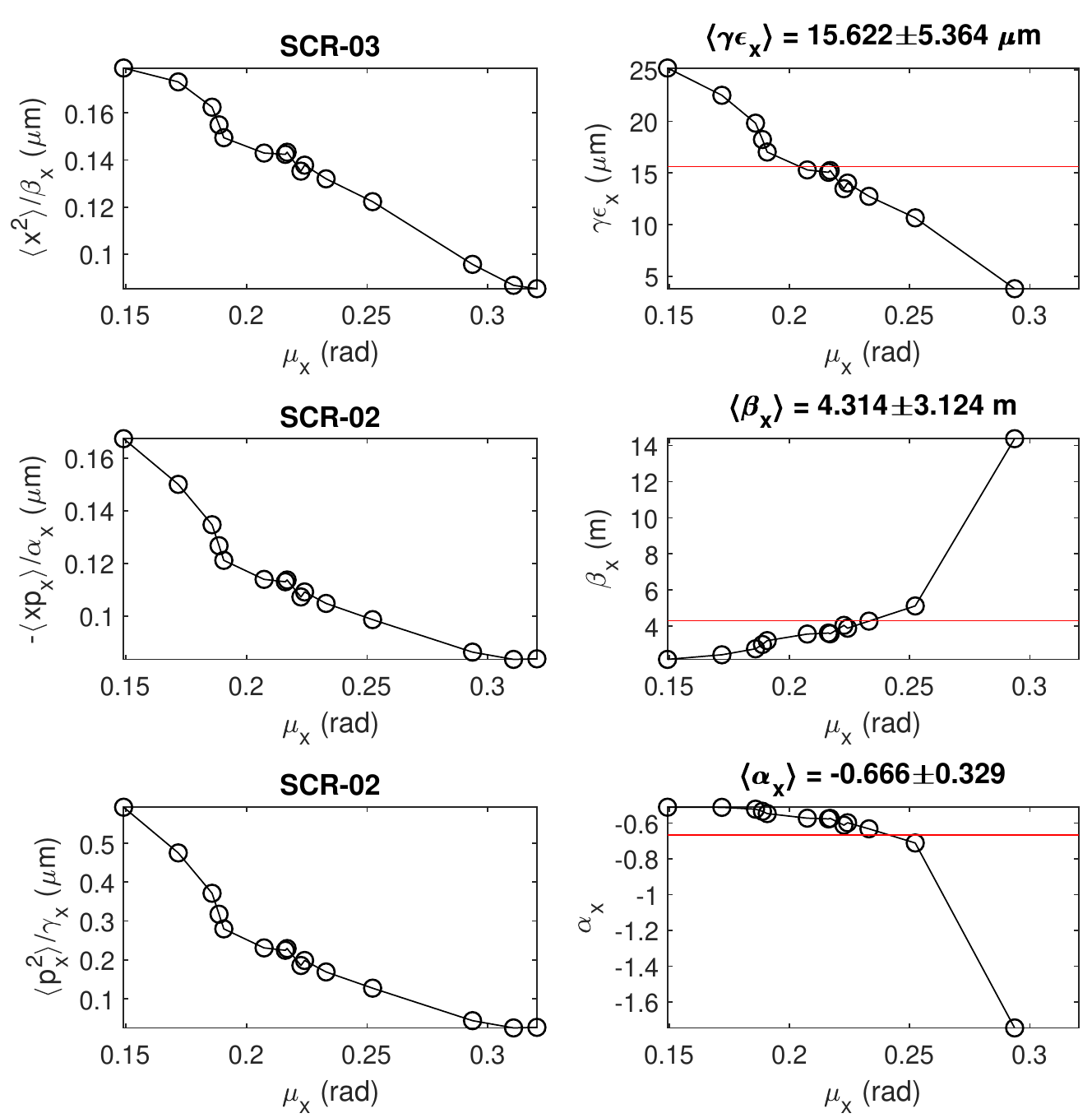}
\caption{Comparison of simulation results for the three-screen optics measurement method using different
phase space distributions, for the transverse horizontal motion.  The plots in the first and third columns show,
for Gaussian and realistic phase space distributions respectively, the scaled covariance matrix elements
$\langle x^2 \rangle / \beta_x$ at SCR-03 (top), $-\langle x p_x \rangle/\alpha_x$ at SCR-02 (middle),
and $\langle p_x^2 \rangle/\gamma_x$ at SCR-02 (bottom).  The plots in the second and fourth columns show,
again for the Gaussian and realistic phase space distributions respectively, the normalised emittance (top),
beta function (middle) and alpha function (bottom).  Note the differences in vertical scales for the sets of plots
for the Gaussian and the realistic phase space distributions.  In the Gaussian case, there is only a very small
variation in the quantities shown, as a function of the quadrupole strengths (used to vary the horizontal phase
advance, given on the horizontal axis on each plot).  The Gaussian distribution was constructed with nominal
parameter values $\gamma\epsilon_x = 4.1\,\mu$m, $\beta_x = 6.49\,$m and $\alpha_x = -1.32$ at SCR-02.
The case of the realistic phase space distribution shows very much larger variation in the calculated covariance
matrix, emittance and optical functions, and displays some similarities with the experimental case
(Fig.~\ref{figthreescreenanalysis}).
 \label{figthreescreenanalysishorizontalsimulation}}
\end{center}
\end{figure*}

\begin{figure*}[th]
\begin{center}
\includegraphics[width=\columnwidth,trim={3.0cm 13.6cm 4.5cm 2.3cm},clip]{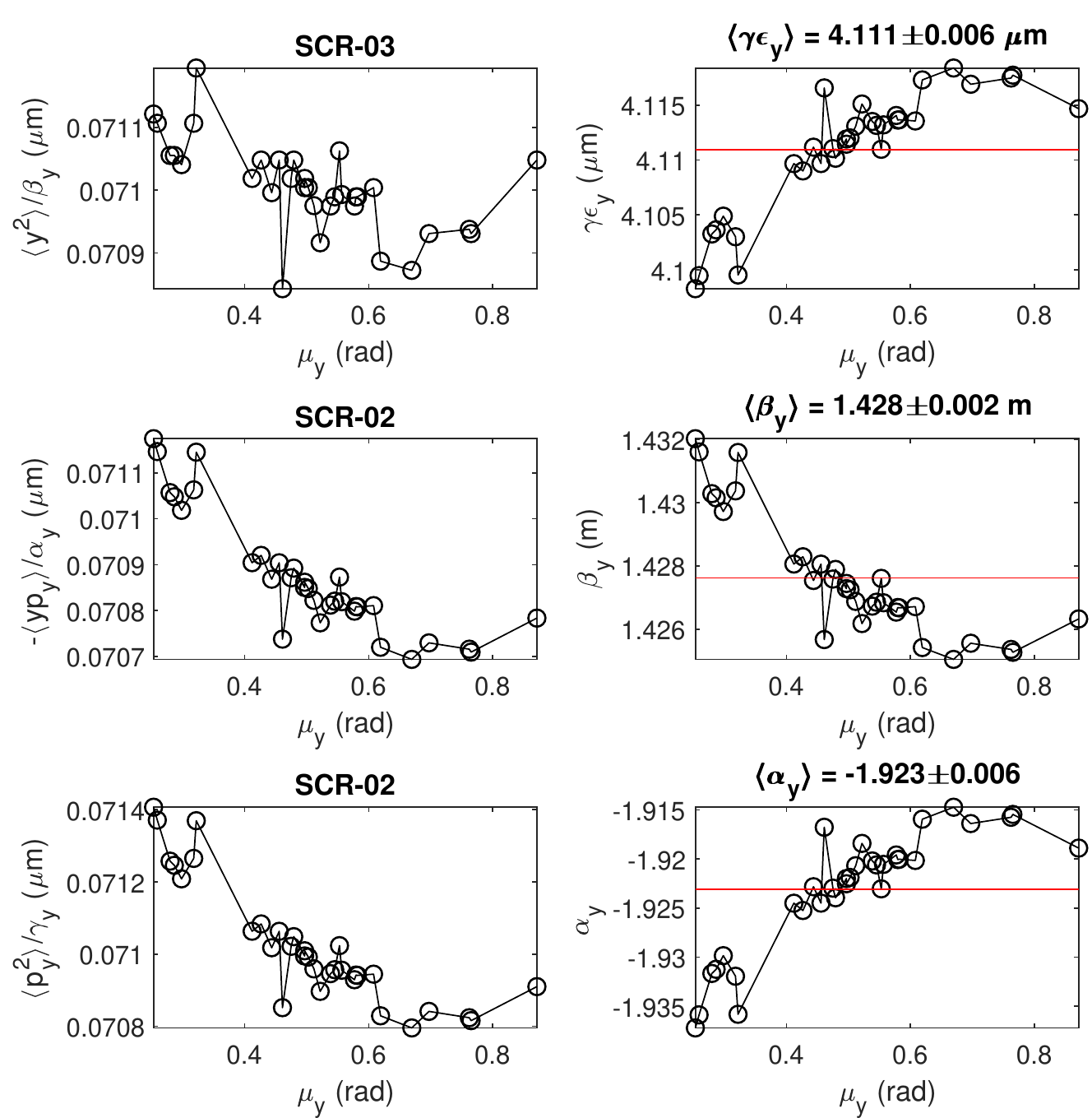}
\includegraphics[width=\columnwidth,trim={3.0cm 13.6cm 4.5cm 2.3cm},clip]{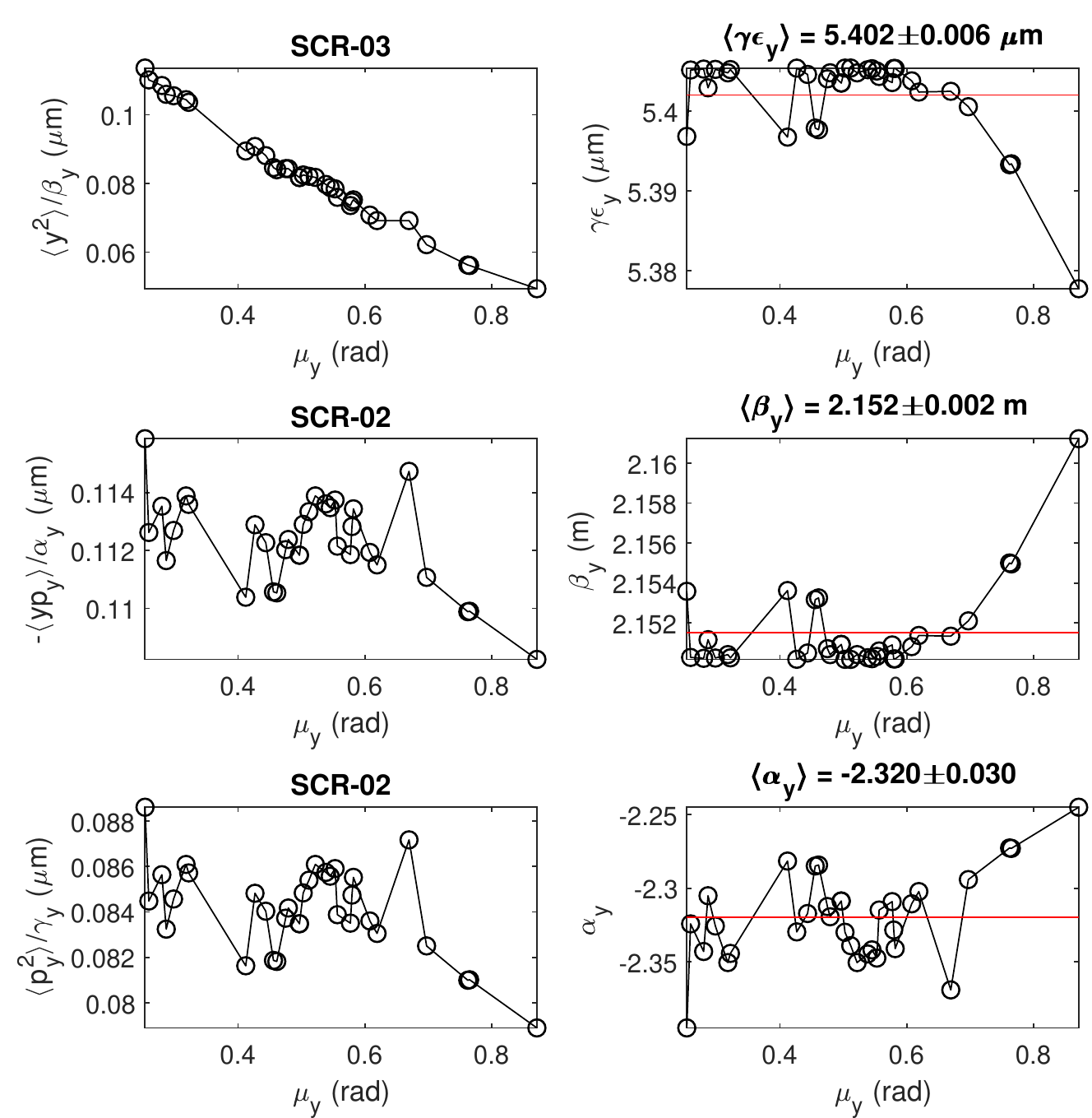}
\caption{Comparison of simulation results for the three-screen optics measurement method using different
phase space distributions, for the vertical motion.  The plots in the first and third columns show,
for Gaussian and realistic phase space distributions respectively, the scaled covariance matrix elements
$\langle y^2 \rangle / \beta_y$ at SCR-03 (top), $-\langle y p_y \rangle/\alpha_y$ at SCR-02 (middle),
and $\langle p_y^2 \rangle/\gamma_y$ at SCR-02 (bottom).  The plots in the second and fourth columns show,
again for the Gaussian and realistic phase space distributions respectively, the normalised emittance (top),
beta function (middle) and alpha function (bottom).  Note the differences in vertical scales for the sets of plots
for the Gaussian and the realistic phase space distributions.  In the Gaussian case, there is only a very small
variation in the quantities shown, as a function of the quadrupole strengths (used to vary the vertical phase
advance, given on the horizontal axis on each plot).  The Gaussian distribution was constructed with nominal
parameter values $\gamma\epsilon_y = 4.1\,\mu$m, $\beta_y = 1.43\,$m and $\alpha_y = -1.92$ at SCR-02.
The case of the realistic phase space distribution shows very much larger variation in the calculated covariance
matrix, emittance and optical functions, and displays some similarities with the experimental case
(Fig.~\ref{figthreescreenanalysis}).
 \label{figthreescreenanalysisverticalsimulation}}
\end{center}
\end{figure*}

We performed simulations to validate the argument that with the three-screen analysis method, the variation
in beam parameters for different quadrupole settings arises from the structure of the phase space distribution.
In the simulations, we created a set of particles with a given phase space distribution, and tracked the particles
(in a computer model of the beamline) from SCR-01 to SCR-02 (the Reconstruction Point) and then to
SCR-03.  At each screen, the horizontal and vertical rms beam size are calculated, and used the same procedure
that was applied to the experimental data to
calculate the covariance matrix at SCR-02, and the emittance and optical parameters at this point.  The tracking
and optical calculations are repeated for different strengths of the quadrupoles, corresponding to those used
in the experiment.  Results for the transverse horizontal plane are shown in
Fig.~\ref{figthreescreenanalysishorizontalsimulation}.  For a Gaussian distribution in phase space, there are
only very small variations in the calculated covariance matrix at SCR-02 and in the optical functions, for 
different quadrupole strengths (the small variations arise from statistical variation in the distribution, resulting
from tracking a finite number of particles).  The simulation can be repeated, but using instead of a Gaussian
distribution, a phase space distribution based on the one found from the tomography study (presented in
Section~\ref{sec:tomography}).  In this case, we see much larger variations in the covariance matrix at
SCR-02 and in the emittance and optical functions at this point, depending on the strengths of the quadrupoles
between SCR-02 and SCR-03.  For some quadrupole strengths, the calculated covariance matrix is unphysical,
and it is not possible to find real values for the emittance or optical functions.  The overall behaviour is qualitatively
similar in some respects with that seen in the experiment (Fig.~\ref{figthreescreenanalysis}).  Results of simulations
for the vertical plane (Fig.~\ref{figthreescreenanalysisverticalsimulation}) again show almost no variation in the
emittance or optical functions as a function of quadrupole
strength for a Gaussian phase space distribution, but behaviour similar to that observed in experiment in the
case of a more realistic phase space distribution based on the results of the tomography analysis.


\subsection{\label{sec:quadscan}Quadrupole scan method}



One of the limitations of the three-screen analysis method described in Section~\ref{sec:threescreen}
is the inability to provide information on beam coupling.  This can be overcome, however, by
combining observations of the transverse beam size at different screens for various strengths of the
quadrupoles between the screens.  If a sufficient number of quadrupole strengths are used, then 
beam size measurements at a single screen provide sufficient data to calculate the $4\times 4$
transverse beam covariance matrix at a point upstream of the quadrupoles.  The $4\times 4$ covariance matrix
has ten independent elements: in principle, just four sets of quadrupole strengths provide twelve beam
size measurements (values for $\langle x^2 \rangle$, $\langle y^2 \rangle$ and $\langle xy \rangle$
for each set of quadrupole strengths), and are more than sufficient to determine the covariance matrix.
In practice, it is desirable to use a greater number of quadrupole strength settings, to over-constrain the covariance matrix.

The quadrupole scan technique that we use is similar to that presented by Prat and Aiba \cite{prat2014}.
The theory can be developed as follows.  The covariance matrix $\Sigma_3$ at a location $s_3$ in the beamline
(SCR-03 in the case of CLARA FE) is related to the covariance matrix $\Sigma_2$ at a location $s_2$
(SCR-02 in CLARA FE) through Eq.~(\ref{eqnsigma3}), where all matrices are now $4\times 4$.  The
relationship between the observable quantities at $s_3$ (assuming a YAG screen at that location) and
the independent elements of $\Sigma_2$ can be written:
\begin{equation}
\left( \begin{array}{c}
\langle x_3^2 \rangle_{(1)} \\
\langle x_3 y_3 \rangle_{(1)} \\
\langle y_3^2 \rangle_{(1)} \\
\langle x_3^2 \rangle_{(2)} \\
\langle x_3 y_3 \rangle_{(2)} \\
\langle y_3^2 \rangle_{(2)} \\
\langle x_3^2 \rangle_{(3)} \\
\langle x_3 y_3 \rangle_{(3)} \\
\langle y_3^2 \rangle_{(3)} \\
\vdots
\end{array} \right) = 
D
\left( \begin{array}{c}
\langle x_2^2 \rangle \\
\langle x_2 p_{x2} \rangle \\
\langle x_2 y_2 \rangle \\
\langle x_2 p_{y2} \rangle \\
\langle p_{x2}^2 \rangle \\
\langle p_{x2} y_2 \rangle \\
\langle p_{x2} p_{y2} \rangle \\
\langle y_2^2 \rangle \\
\langle y_2 p_{y2} \rangle \\
\langle p_{y2}^2 \rangle
\end{array} \right),
\label{beamsizetosigma2dof}
\end{equation}
where $\langle x_3^2 \rangle_{(n)}$ represents the mean square horizontal transverse beam size measured
at $s_3$ for a particular set of quadrupole strengths, and similarly for $\langle y_3^2 \rangle_{(n)}$ and
$\langle x_3 y_3 \rangle_{(n)}$.  With  measurements of the beam distribution in co-ordinate space at $s_3$
for $N$ different sets of quadrupole strengths, $D$ is a $3N\times 10$ matrix.  The elements of $D$ can be
found, using Eq.~(\ref{eqnsigma3}), from the transfer matrices $M_{23}$ from $s_2$ to $s_3$ (with each set
of three rows in $D$ corresponding to a single set of quadrupole strengths).  Explicit expressions for the
elements of $D$ (for a given transfer matrix) are as follows:
\begin{widetext}
\begin{equation}
D^\mathrm{T} = \left( \begin{array}{ccc}
m_{1,1}^2 & m_{1,1}m_{3,1} & m_{3,1}^2 \\
2m_{1,1}m_{1,2} & m_{1,2}m_{3,1}+m_{1,1}m_{3,2} & 2m_{3,1}m_{3,2}  \\
2m_{1,1}m_{1,3} & m_{1,3}m_{3,1}+m_{1,1}m_{3,3} & 2m_{3,1}m_{3,3}  \\
2m_{1,1}m_{1,4} & m_{1,4}m_{3,1}+m_{1,1}m_{3,4} & 2m_{3,1}m_{3,4}  \\
m_{1,2}^2 & m_{1,2}m_{3,2} & m_{3,2}^2  \\
2m_{1,2}m_{1,3} & m_{1,3}m_{3,2}+m_{1,2}m_{3,3} & 2m_{3,2}m_{3,3}  \\
2m_{1,2}m_{1,4} & m_{1,4}m_{3,2}+m_{1,2}m_{3,4} & 2m_{3,2}m_{3,4}  \\
m_{1,3}^2 & m_{1,3}m_{3,3} & m_{3,3}^2  \\
2m_{1,3}m_{1,4} & m_{1,4}m_{3,3}+m_{1,3}m_{3,4} & 2m_{3,3}m_{3,4}  \\
m_{1,4}^2 & m_{1,4}m_{3,4} & m_{3,4}^2 
\end{array} \right),
\end{equation}
\end{widetext}
where $m_{i,j}$ is the $(i,j)$ element of the transfer matrix $M_{23}$ (for a given set of quadrupole strengths).
Given observations
of the beam profile at $s_3$ for a number of different sets of quadrupole strengths, and the corresponding
values for the elements of $D$, the elements of the covariance matrix at $s_2$ may be found by inverting
Eq.~(\ref{beamsizetosigma2dof}).  Since $D$ is not a square matrix, the pseudo-inverse of $D$ (found,
for example, using singular value decomposition) must be used instead of the strict inverse.

It is worth noting that whereas in one degree of freedom it is possible to obtain the elements of the
covariance matrix at the Reconstruction Point by varying the strength of a single quadrupole between
the Reconstruction Point and the Observation Point, this is not the case in two degrees of freedom.
To understand the reason for this, consider the case of a single thin quadrupole with the Reconstruction
Point $s_2$ at the upstream (entrance) face of the quadrupole, and the Observation Point $s_3$ some distance
downstream from the quadrupole.  The elements of the covariance matrix $\langle x_3^2 \rangle$,
$\langle x_3 y_3 \rangle$ and $\langle y_3^2 \rangle$ each have a quadratic dependence on the
quadrupole strength, with coefficients determined by the elements of the covariance matrix at the
Reconstruction Point.  By fitting the quadratic curves obtained from a quadrupole scan we therefore
obtain nine constraints (three for each of the observed elements of the covariance matrix at $s_2$);
however, the covariance matrix at $s_2$ has ten independent elements (in two degrees of freedom).
The problem is therefore underconstrained: in the context of Eq.~(\ref{beamsizetosigma2dof}) this
is manifest as the matrix $D$ having fewer non-zero singular values than are required to determine
uniquely the elements of the covariance matrix at the Observation Point.  Although it is always possible
to ``invert'' $D$ using singular value decomposition, the procedure in this case would yield a solution
for the covariance matrix that minimises the sum of the squares of the matrix elements: there is no
reason to suppose that this least-squares matrix is near the correct solution.  To address this
problem, however, it is only necessary to collect data from a scan of two quadrupoles at different
locations between the Observation Point and the Reconstruction Point.  This breaks the degeneracy
in the system, and (if the system is properly designed) more than ten singular values of $D$ will be non-zero:
in other words, the system becomes over-constrained, rather than under-constrained.

The same data collected for the three-screen method can be used in the analysis using the
quadrupole scan method, and the same practical considerations (concerning, for example, the desirability
of a beam waist at the Reconstruction Point, and maintaining an approximately round beam at the
Observation Point) apply.  However, it should be noted
that for the three-screen method, the observed beam sizes at all three screens are used to reconstruct
the covariance matrix: an independent reconstruction is obtained for each point in the quadrupole scan.
For the quadrupole scan analysis method, on the other hand, we use only the observed beam size at
a single screen (SCR-03 in this case) and combine all the measurements for different quadrupole strengths
to calculate the elements of the covariance matrix.  In effect, we calculate the size and shape of the distribution
in phase space based on the widths of projections at many different phase angles: this leads to a more
reliable result than is obtained using the three-screen analysis method, for which only three different phase
angles are used.  Nevertheless, even for a large number of phase angles, the quadrupole scan method does
not provide the same detailed information on the phase space distribution that is provided by the tomography
method (discussed in Section~\ref{sec:tomography}).  Rather, it fits a phase space distribution that may have
significant detailed structure with a simple Gaussian distribution.

\begin{figure}[th]
\begin{center}
\includegraphics[width=0.95\columnwidth,trim={3.0cm 13.0cm 4.5cm 2.0cm},clip]{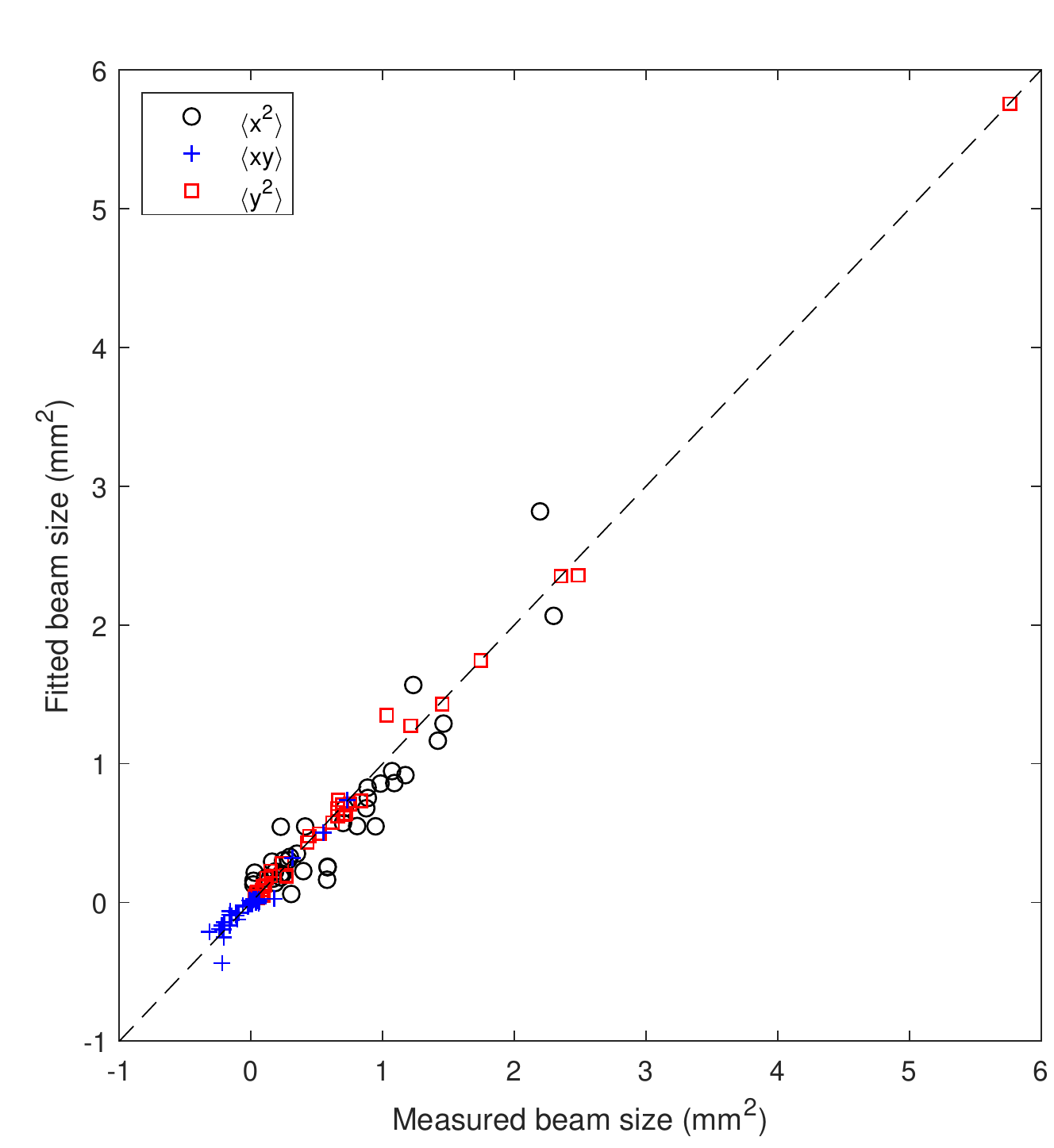}
\caption{Beam size at SCR-03 reconstructed from a quadrupole scan analysis compared with the beam
size observed directly on the screen.  Each point corresponds to a different set of quadrupole strengths.
Circles indicate the horizontal beam size $\langle x^2 \rangle$; boxes indicate the vertical beam size
$\langle y^2 \rangle$; crosses indicate the horizontal-vertical correlation $\langle xy \rangle$.  The distances
of the points from the dashed line (which passes through the origin with unit gradient) indicates the residuals
to the fit.
\label{figquadscananalysis}}
\end{center}
\end{figure}

Figure \ref{figquadscananalysis} shows the residuals from a fit based on data from a quadrupole scan
in CLARA FE made with nominal machine settings.  Each point indicates the
observed and fitted beam size ($\langle x^2 \rangle$, $\langle y^2 \rangle$ or $\langle xy \rangle$) at the
Observation Point for a different set of strengths of the quadrupoles between SCR-02 (the Reconstruction
Point) and SCR-03 (the Observation Point).  The results may also be validated
by comparing the beam size predicted at the Reconstruction Point with the actual beam size
observed at this point.  For the case shown, the agreement is within about 15\%.

\subsection{\label{sec:tomography}Phase space tomography}



Finally, it is possible to use phase space tomography to
construct a more detailed representation of the beam properties than is provided by just the emittance
and optical functions. In principle, the tomography method is similar to the quadrupole
scan, in that by observing the beam image on a screen for different strengths of a set of upstream
quadrupoles, it is possible to reconstruct the phase space distribution at a point upstream of the
quadrupoles.  The difference is that for the quadrupole scan, only the rms beam sizes are used in
the analysis: tomography uses all the information from the (observed) beam distribution to produce
a more detailed reconstruction of the phase space distribution of the beam.  When tomography is carried
out in co-ordinate space, two-dimensional images (projections) on a screen for different orientations of
an object are used to reconstruct a three-dimensional representation of the object.  In phase space
tomography, different ``orientations'' correspond to rotations in phase space, which are achieved by
changing the horizontal or vertical phase advances between the Reconstruction Point and the Observation
Point.  Mathematically, the analysis is essentially the same as in co-ordinate space, and standard algorithms
developed for tomography in co-ordinate space (such as filtered back-projection, or maximum entropy
\cite{kakslaney2001, minerbo1979, mottershead1985}) can be applied to phase space tomography.

For analysis of data from CLARA FE, we have used a form of algebraic reconstruction.
The procedure may be outlined as follows.  For simplicity we consider just a single degree of freedom:
the generalisation to two (or more) degrees of freedom is straightforward.  Let $\psi$ be a vector in
which each element $\psi_j$ represents the beam density at a particular point $(x_j, p_{xj})$ in phase
space at the Reconstruction Point. Assuming that the points are evenly distributed on a grid in phase
space, then the projected beam density at the Observation Point, $\rho^{(1)}_i$ at a point $x = x_i$
in co-ordinate space, can be  be written as a matrix multiplication:
\begin{equation}
\rho^{(1)}_i = \sum_j P^{(1)}_{ij} \psi_j,
\label{tomgraphyartequation}
\end{equation}
where the matrix $P^{(1)}$ has elements:
\begin{equation}
P^{(1)}_{ij} = \left\{ \begin{array}{l}
1 \textrm{ if } x_i = m^{(1)}_{1,1} x_j + m^{(1)}_{1,2} p_{xj}, \\
0 \textrm{ otherwise.}
\end{array} \right.
\end{equation}
$m^{(1)}_{1,1}$ and $m^{(1)}_{1,2}$ are elements of the transfer matrix from the Reconstruction Point to the
Observation Point, for a given set of quadrupole strengths.  If the vector $\rho$ has $\mathcal{N}$ elements
$\rho_i$, and there are $\mathcal{N}^2$ points $(x_j, p_{xj})$ in phase space , then $P^{(1)}$ is an
$\mathcal{N}\times \mathcal{N}^2$ matrix.  If the transfer matrix from the Reconstruction
Point to the Observation Point is changed (e.g.~by changing the strengths of the quadrupoles between
the two points), then we construct a new vector $\rho^{(2)}$ from the new image at the Observation Point,
corresponding to the new transfer matrix.  In general, for the $n$th transfer matrix, we have:
\begin{equation}
\rho^{(n)}_i = \sum_j P^{(n)}_{ij} \psi_j.
\label{psprojection}
\end{equation}
Note that the phase space density $\psi$ is constant, because $\psi$ refers to a point upstream of any
quadrupoles whose strength is changed during the measurements.  We can combine the observations simply
by stacking the vectors $\rho^{(n)}$ and the matrices $P^{(n)}$:
\begin{equation}
\rho = \left( \begin{array}{c}
\rho^{(1)} \\
\rho^{(2)} \\
\vdots \\
\rho^{(n)} \end{array} \right),
\quad \textrm{ and } \quad
P_{ij} = \left( \begin{array}{c}
P^{(1)} \\
P^{(2)} \\
\vdots \\
P^{(n)} \end{array} \right).
\end{equation}
$\rho$ is a vector with $n\mathcal{N}$ elements, and $P$ is an $n\mathcal{N}\times \mathcal{N}^2$ matrix.
In terms of the pseudo-inverse $P^\dagger$ of $P$, we have the following formula for the phase space density
at the Reconstruction Point:
\begin{equation}
\psi_j = \sum_i P^\dagger_{ji}\rho_i .
\label{eqntomography4d}
\end{equation}

We perform the analysis in normalised phase space \cite{hock2011}, in which the phase space
variables $(x_\mathrm{N}, p_{x\mathrm{N}})$ are defined by:
\begin{equation}
\left( \begin{array}{c}
x_\mathrm{N} \\
p_{x\mathrm{N}}
\end{array} \right) = 
\left( \begin{array}{cc}
\frac{1}{\sqrt{\beta_x}} & 0  \\
\frac{\alpha_x}{\sqrt{\beta_x}} & \sqrt{\beta_x} 
\end{array} \right)
\left( \begin{array}{c}
x \\
p_x 
\end{array} \right),
\end{equation}
where $\alpha_x$ and $\beta_x$ are the Courant--Snyder functions at the given point in the beam line.
The transfer matrix in normalised phase space between any two points in the beam line is represented
by a pure rotation matrix, with rotation angle given by the phase advance.  This simplifies the implementation
of the algebraic tomography method described above.  A further advantage of working in normalised phase space is
that if the Courant--Snyder functions at the Reconstruction Point are chosen to match the beam distribution,
then the beam distribution in phase space at this point will be perfectly circular: this improves the accuracy
with which parameters such as the emittance may be calculated.  Note that, since we do not
know in advance the actual Courant--Snyder parameters describing the beam distribution at the
Reconstruction Point, we need to make some estimate based on (for example) simulations or a quadrupole
scan analysis.  In practice, it is not essential for the estimated parameters to match exactly the actual beam
parameters: any discrepancy will simply lead to an elliptical distortion of the beam distribution in normalised
phase space.  To work in normalised phase space is straightforward: all that is necessary is to scale
the co-ordinate axis for the observed beam projection by a factor $1/\sqrt{\beta^\mathrm{OP}_x}$,
where $\beta^\mathrm{OP}_x$ is the Courant--Snyder beta function at the Observation Point
calculated from the estimated (fixed) Courant--Snyder functions at the Reconstruction Point and
the transfer matrix from the Reconstruction Point to the Observation Point.

Rather than compute the pseudo-inverse of $P$, we solve Eq.~(\ref{psprojection}) iteratively, using a
least-squares method.  For the computation of the phase space in a single degree of freedom, we apply
a constraint that the particle density must be positive at all points in phase space.  However, applying
this constraint carries considerable computation overhead, and for computation of the phase space in
two degrees of freedom, which has considerably greater computational cost than the case of a single
degree of freedom, we do not constrain the least-squares solver in this way.  This can result in negative
(unphysical) values for the particle density at some points in phase space; however, when a good fit is
achieved, the negative values make a relatively small contribution to the overall phase space distribution.

\begin{figure}[th]
\begin{center}
\includegraphics[width=0.99\columnwidth,trim={3.2cm 13.2cm 4.5cm 2.6cm},clip]{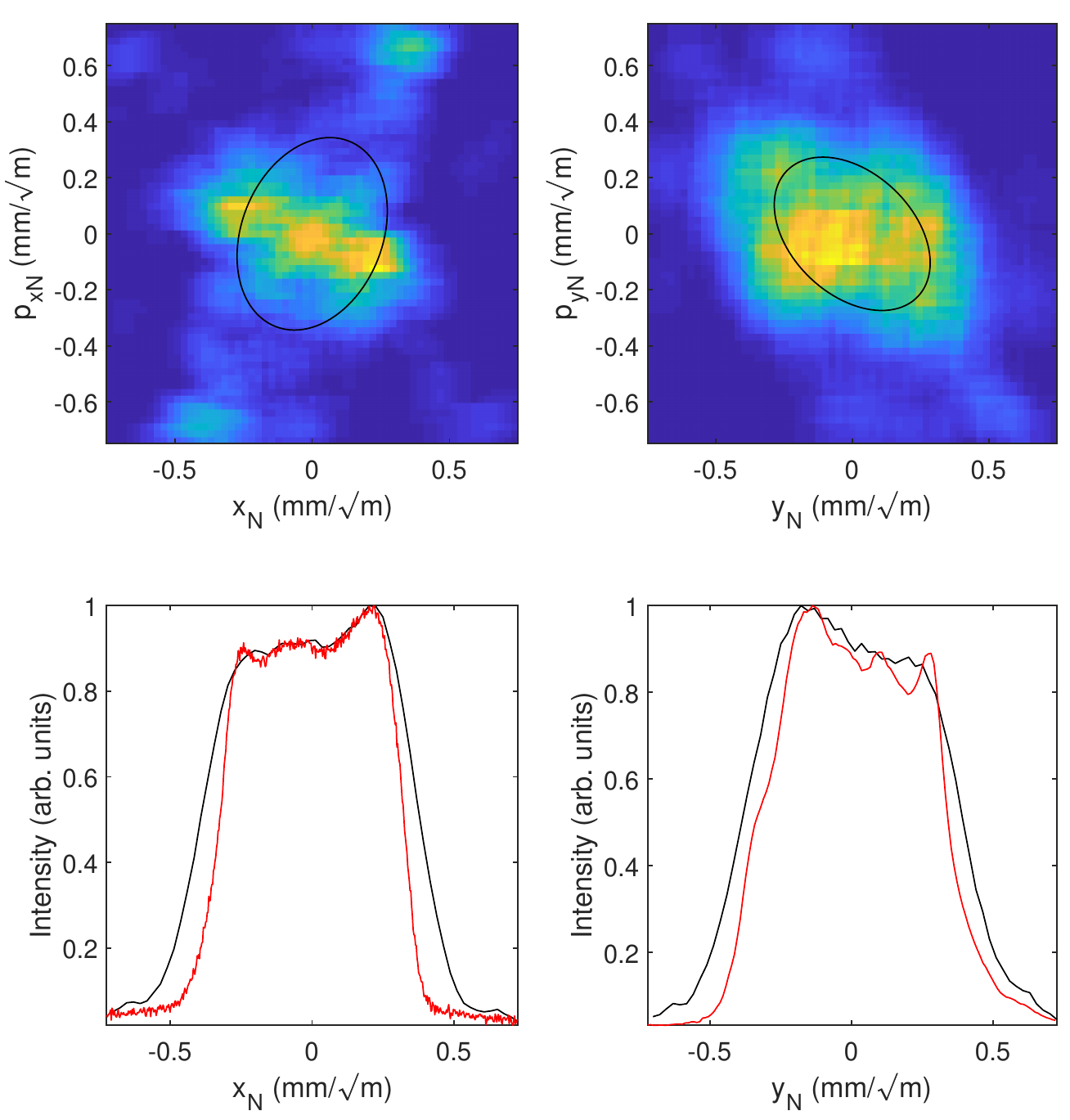}
\caption{Results from phase space tomography in CLARA FE, treating horizontal and vertical
phase spaces separately.  The top plots show the charge density in the beam at SCR-02 in the horizontal (left)
and vertical (right) phase spaces, in normalised variables (co-ordinates scaled by the square root of the
beta function).  The black ellipses show the emittance ellipses fitted to the phase space distribution, so as to
satisfy (\ref{emittanceellipse}).  The bottom plots show
the density projected onto the horizontal (left) or vertical (right) axes: black lines are from the tomographic
reconstruction, red lines are from direct observation of the beam image on SCR-02.
\label{figtomography2d}}
\end{center}
\end{figure}

Although there is no need for the phase advances between the Reconstruction Point and Observation Point
to be evenly distributed over the set of observations for different quadrupole strengths, it generally improves
the accuracy of the tomography analysis to use as wide a range of phase advances as possible, with roughly
uniform spacing: this maximises the overall constraints on the phase space distribution for a given number of
observations.  The sets of quadrupole strengths identified in the preparatory simulations (described above)
were chosen to provide a wide range of phase advances.  The same data (screen images at the Observation
Point, for a range of different quadrupole strengths) can be used for the three-screen analysis (described in
Section~\ref{sec:threescreen}), the quadrupole scan analysis (described in Section~\ref{sec:quadscan}) and
the tomography analysis described here. Figure \ref{figtomography2d} shows a set of results from tomography
analysis for the nominal machine settings, and in which the horizontal and vertical phase spaces are treated
independently. As was the case for the quadrupole scan method, the results may be validated by comparing
the predicted beam size at the Reconstruction Point with the beam observed directly at this point (SCR-02):
the results of the comparison are shown in the lower plots in Fig.~\ref{figtomography2d}.  There is good
agreement, and it can be clearly seen that the tomography analysis reveals some features of the charge
distribution in phase space that are not obtained from the quadrupole scan analysis.

\begin{figure}[th]
\begin{center}
\includegraphics[width=0.99\columnwidth,trim={2.7cm 15.2cm 1.1cm 7.4cm},clip]{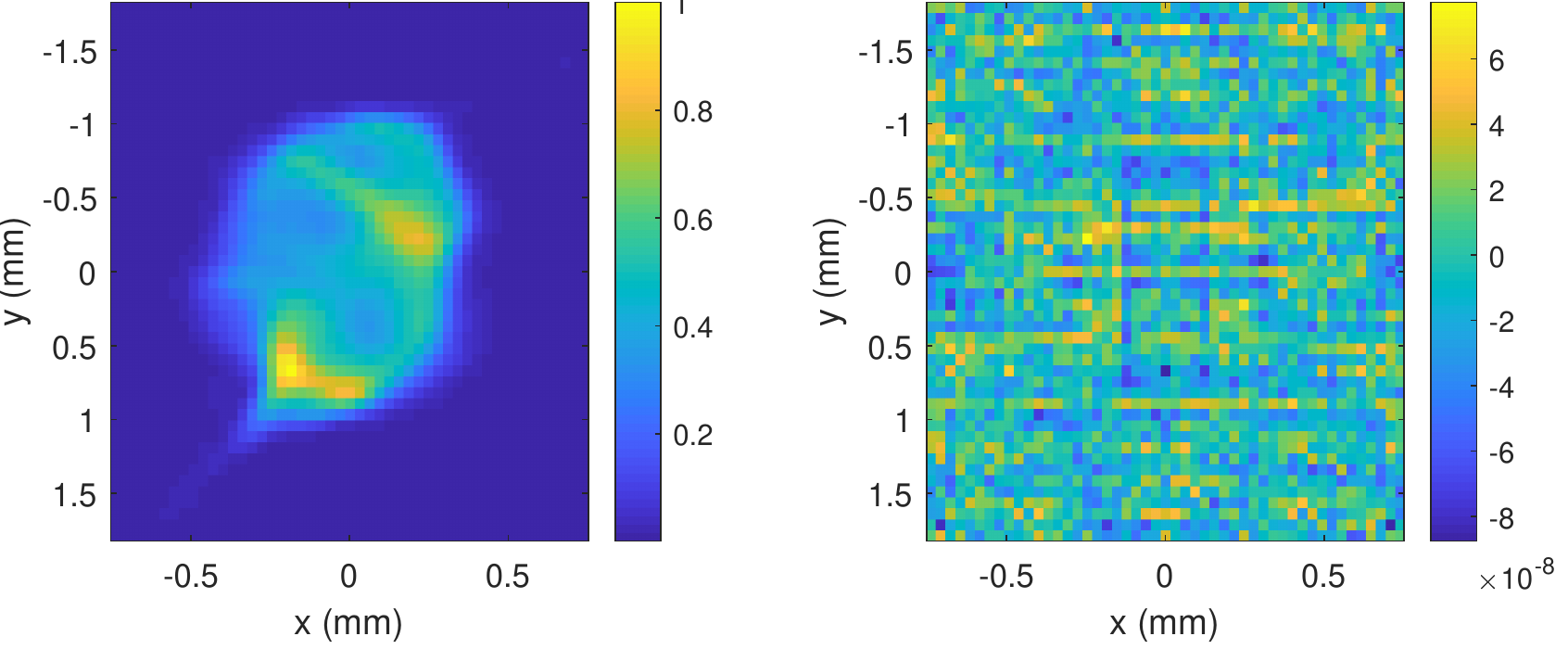}
\caption{Left: typical beam image at SCR-03 (the Observation Point) for one point
in a quadrupole scan.  The image reconstructed from the four-dimensional phase space density found using
Eq.~(\ref{eqntomography4d}) extended to two degrees of freedom (projecting the phase space density
$\psi$ into co-ordinate space) is visually indistinguishable from the observed image.  Right: residuals of the
fit, representing the difference between the intensity of each pixel in the observed image, and the intensity of
the corresponding pixel in the image reconstructed from the four-dimensional phase space density.
The beam image (left) is scaled so that the intensity varies between 0 (dark blue)
and 1 (yellow); on this scale, the largest residuals (right) are of order $10^{-7}$.
\label{figtomography4dresiduals}}
\end{center}
\end{figure}

\begin{figure*}[th]
\begin{center}
\parbox{\columnwidth}{
\includegraphics[width=0.9\columnwidth,trim={3.2cm 4.0cm 4.5cm 2.0cm},clip]{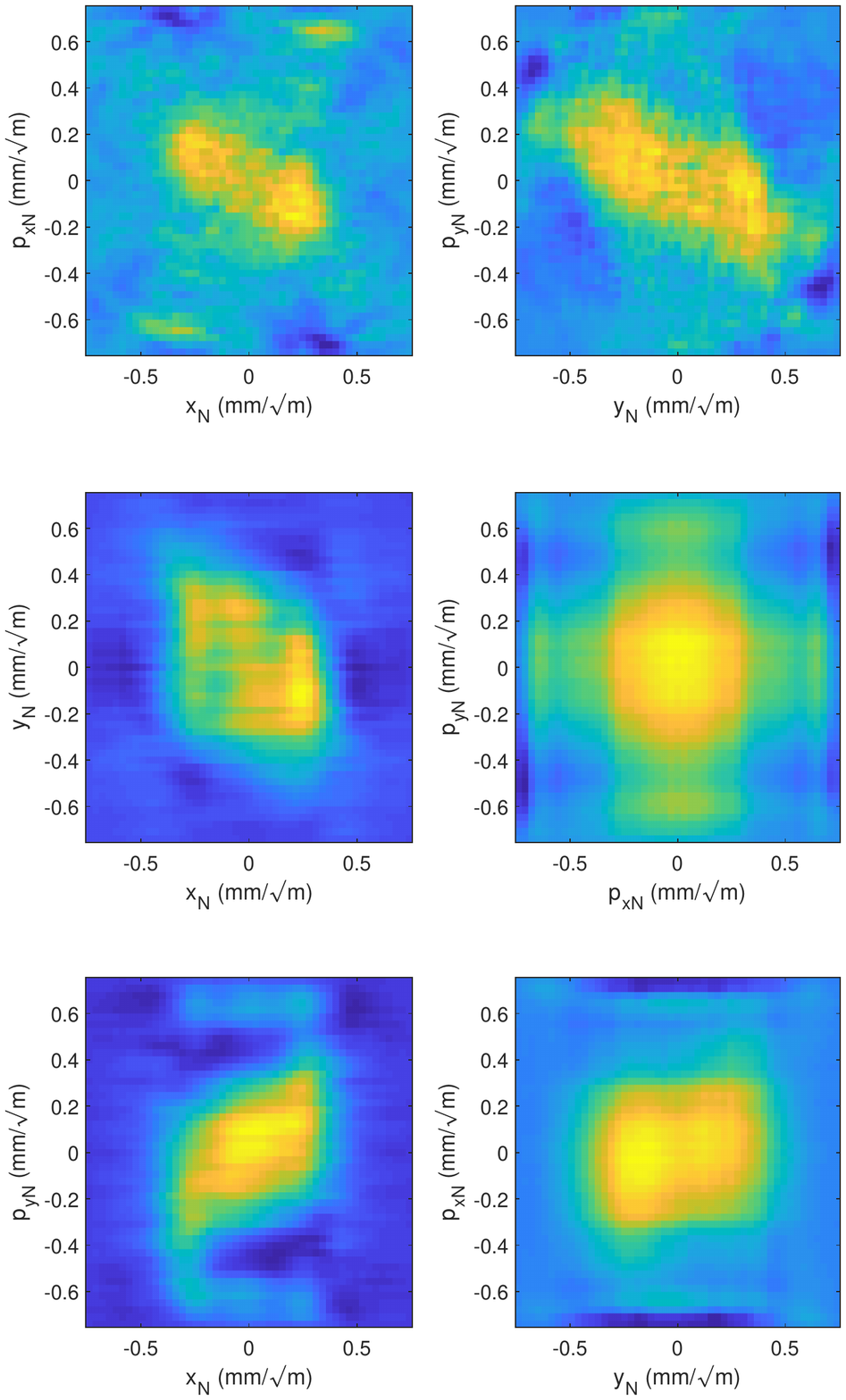} \\
(a) Phase-space distribution from tomography analysis of experimental data.
}
\parbox{\columnwidth}{
\includegraphics[width=0.9\columnwidth,trim={3.2cm 4.0cm 4.5cm 2.0cm},clip]{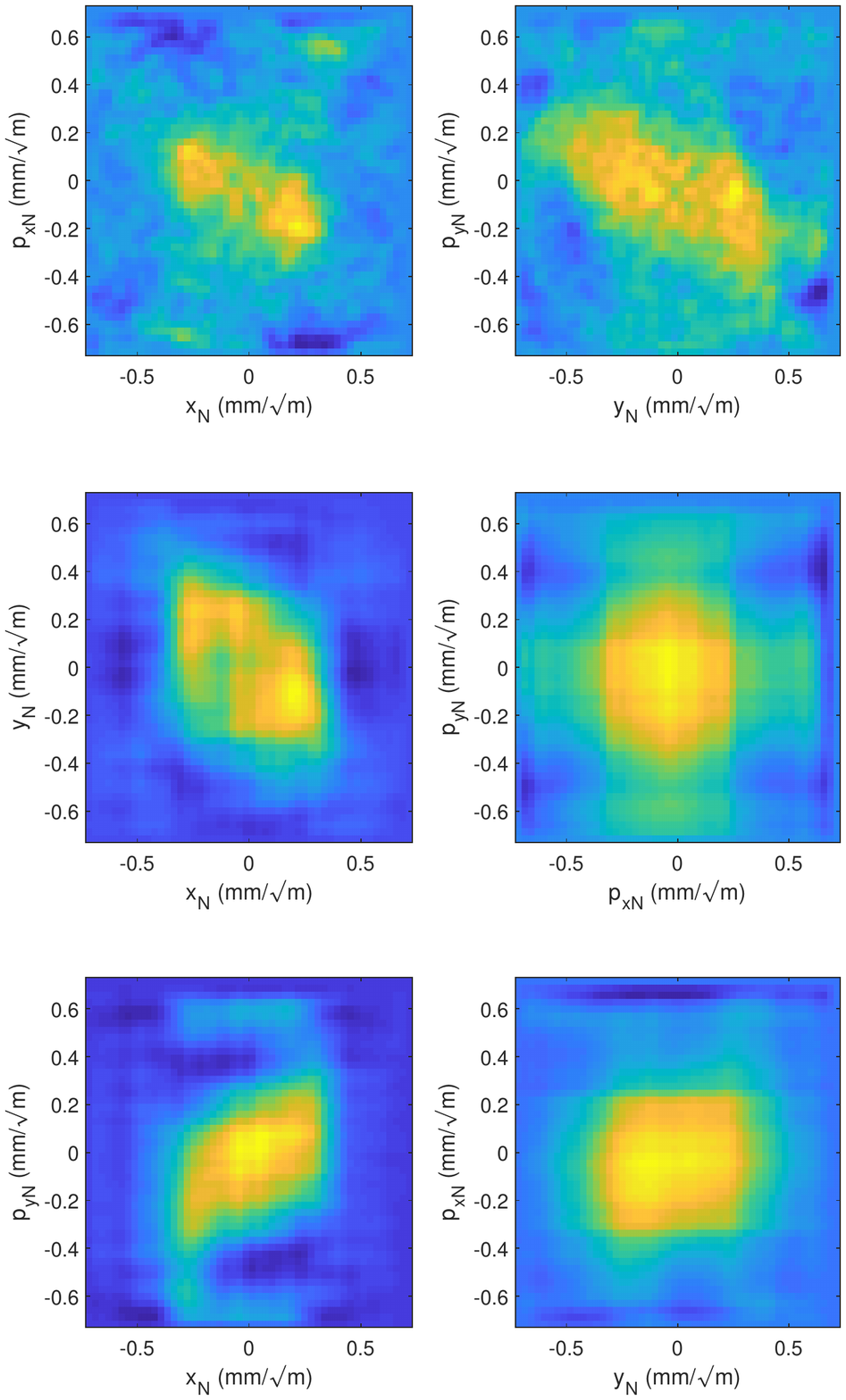} \\
(b) Phase-space distribution from tomography analysis of simulated data.
}
\caption{Projections of beam density in normalised phase space, found from phase space tomography in two degrees of freedom in
CLARA FE.  Each plot shows a different projection of the charge density from four-dimensional phase space, using
normalised phase space variables.
Coupling in the beam is evident in the tilt of the charge distribution in the cases that the
axes refer to different degrees of freedom.  The left-hand set of plots (a) shows the phase space distribution reconstructed
from experimental data; the right-hand set of plots (b) shows the results of the tomography analysis applied to
simulated data based on the measured phase space distribution, to validate the technique.  Although there are some
differences between the analysis results from the experimental data and the results from the simulated data, there
is overall very good agreement in the phase space distribution found in each case, and in the emittances and
optical functions corresponding to fitted emittance ellipses (see Table~\ref{tablebeamparametercomparison}).
\label{figtomography4d}}
\end{center}
\end{figure*}

Treating the horizontal and vertical phase spaces separately in the analysis means that no information is
provided on coupling in the beam, which may arise (for example) from incorrect setting of the bucking coil
at the electron source.  It is possible to extend the tomography analysis from a single degree of freedom, to treating
two degrees of freedom simultaneously \cite{hock2013}.  Applying this technique to the case considered here,
the resulting four-dimensional phase space reconstruction includes information about the betatron coupling
in the beam.  Some results from experimental data (screen images) are shown in Fig.~\ref{figtomography4d}(a).
Generally, the fit using Eq.~(\ref{eqntomography4d}) of the phase space beam density to the observed images
is very good: the residuals from a typical example are shown in Fig.~\ref{figtomography4dresiduals}.

One drawback of applying phase space tomography in two degrees of freedom is that the matrix $P$ in
Eq.~(\ref{tomgraphyartequation}) becomes very large: in one degree of freedom, to reconstruct the phase space
distribution with resolution $\mathcal{N}$ in each dimension using $n$ observations, $P$ will be an
$n\mathcal{N}\times \mathcal{N}^2$ matrix. In two degrees of freedom (four-dimensional phase space),
$P$ will be an $n\mathcal{N}^2\times \mathcal{N}^4$ matrix: even for a relatively coarse reconstruction,
with $\mathcal{N}$ of order 50, computing $P$ and applying its inverse can require significant computational
resources.  The situation is eased somewhat by the fact that in practice, $P$ is a sparse matrix, and this
allows a significant reduction in the computer memory that would otherwise be required; nevertheless, the
required computational resources can be a limit on the resolution with which the phase space in two degrees
of freedom may be reconstructed.  The results shown here use a four-dimensional phase space resolution
$\mathcal{N} = 69$. 

Projections from the four-dimensional phase space density found from experimental data in CLARA FE
are shown in Fig.~\ref{figtomography4d}\,(a).  To validate the technique, we take the four-dimensional
phase space distribution, and use it in a simulation to construct a set of images on SCR-03 corresponding to
different quadrupole strengths.  We then take the simulated images, and again apply the tomography analysis:
the results are shown in Fig.~\ref{figtomography4d}\,(b).  Although there are some differences between
the original and reconstructed distributions they are sufficiently close to indicate that the technique
potentially has good accuracy.  We also find that there is good agreement between the emittances and
optics functions obtained by fitting ellipses to the projections of the phase space into the horizontal and
vertical planes (see Table~\ref{tablebeamparametercomparison}).

\begin{figure*}[th]
\begin{center}
\begin{tabular}{cc}
\includegraphics[width=0.48\textwidth,trim={3cm 13cm 3cm 2.5cm},clip]{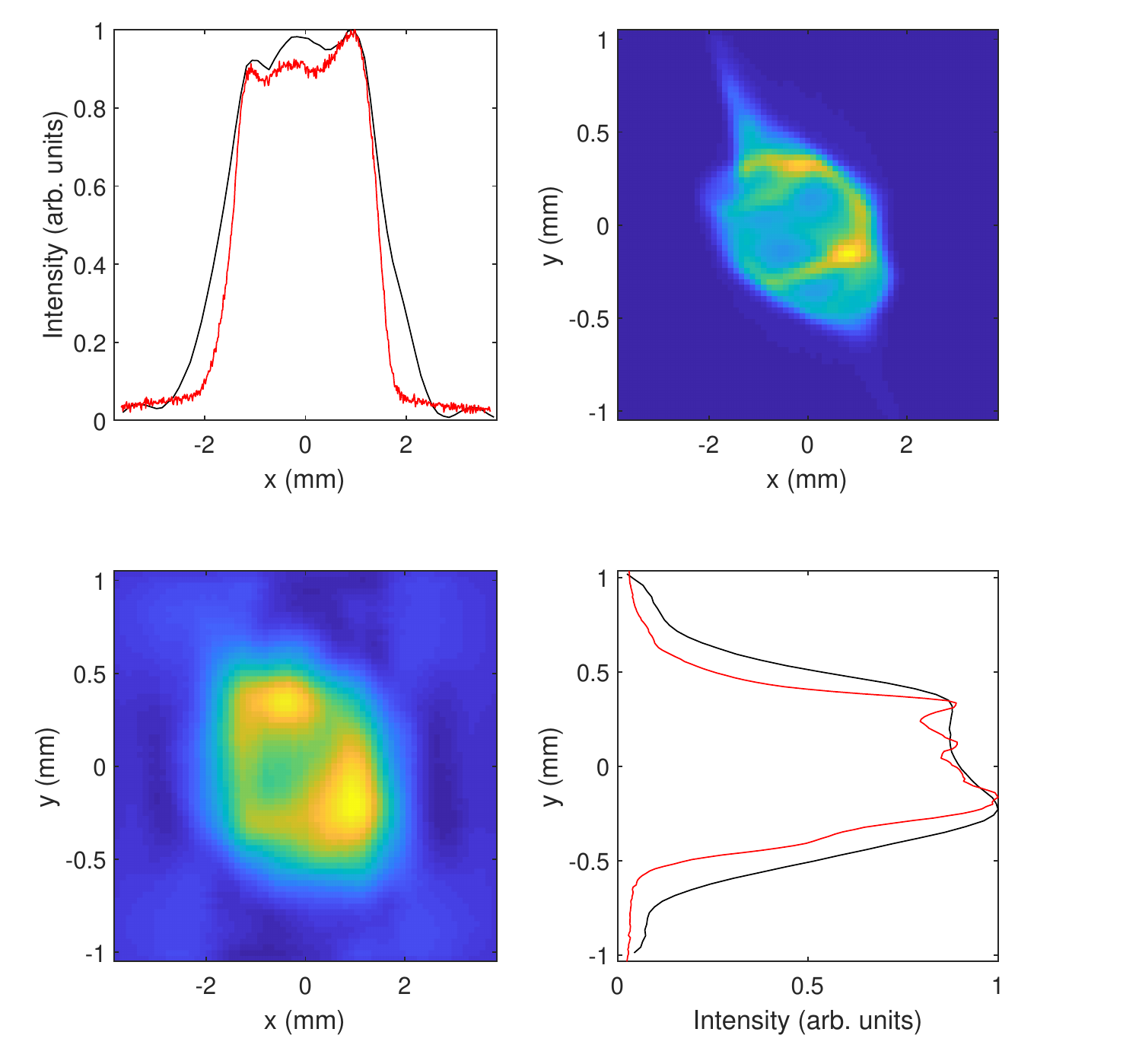} &
\includegraphics[width=0.48\textwidth,trim={3cm 13cm 3cm 2.5cm},clip]{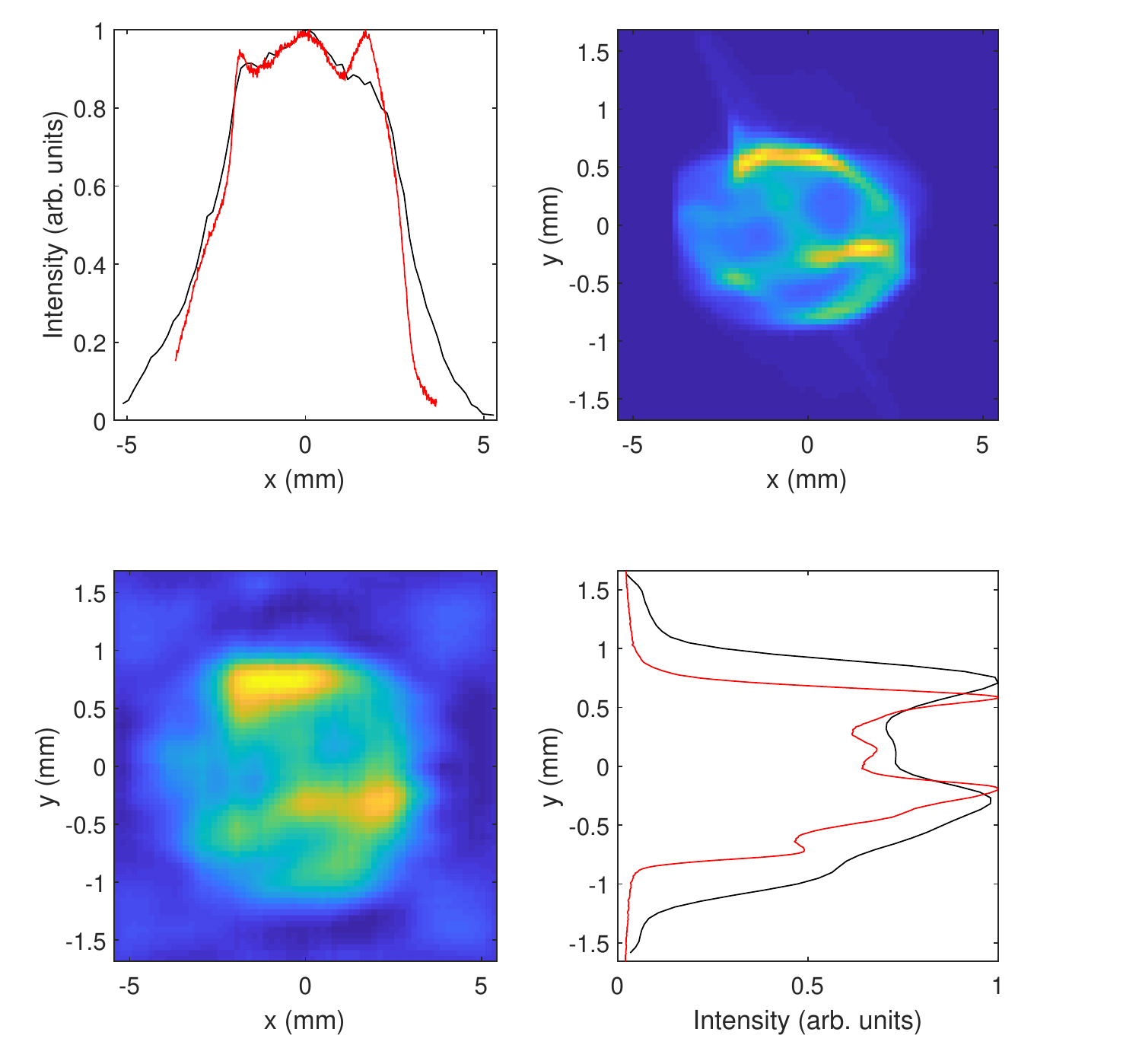} \\
(a) Main solenoid -125\,A, bucking coil -2.2\,A, &
(b) Main solenoid -125\,A, bucking coil -2.2\,A, \\
bunch charge 10\,pC &
bunch charge 20\,pC \\
 & \\
\includegraphics[width=0.48\textwidth,trim={3cm 13cm 3cm 2.5cm},clip]{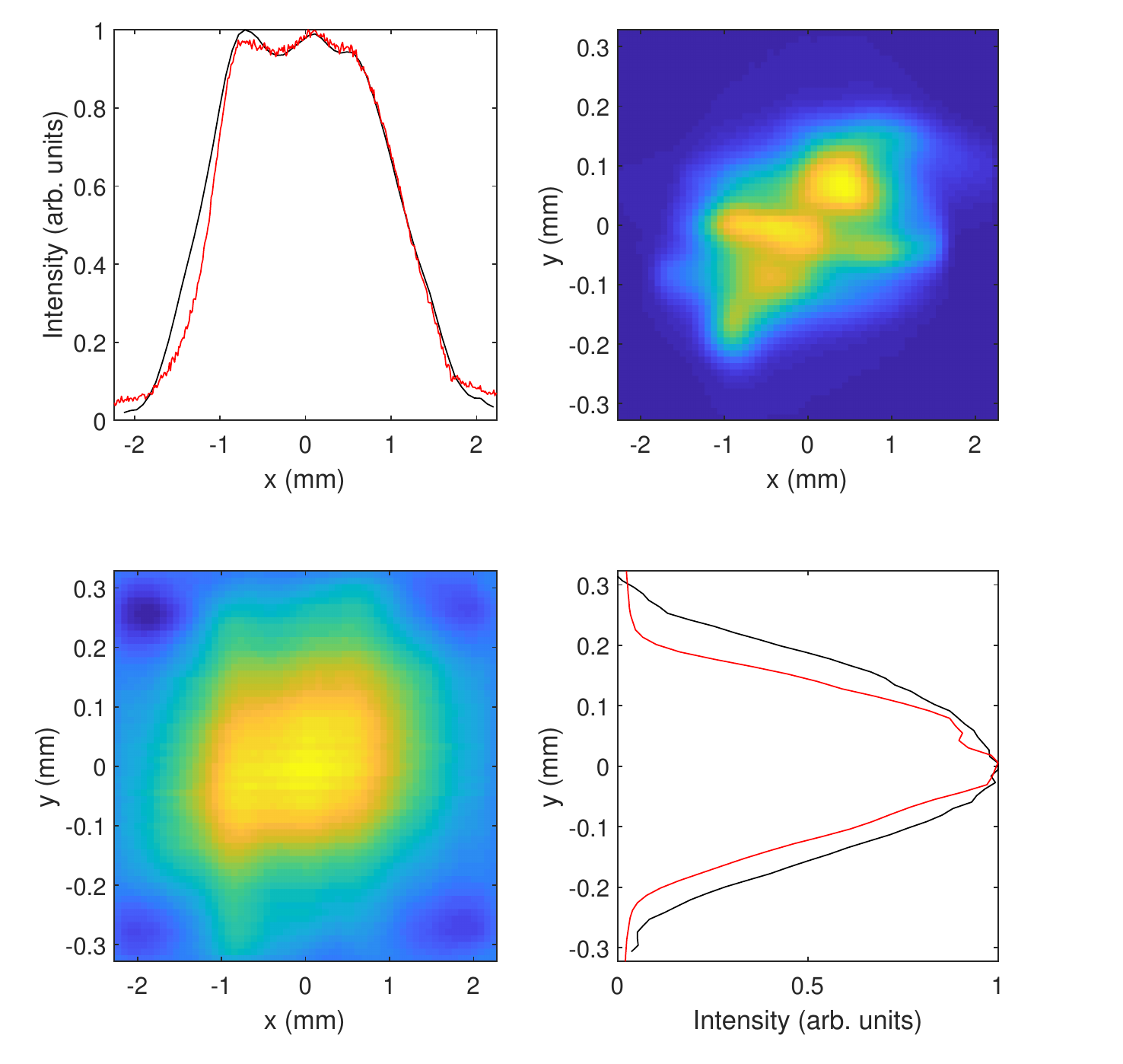} &
\includegraphics[width=0.48\textwidth,trim={3cm 13cm 3cm 2.5cm},clip]{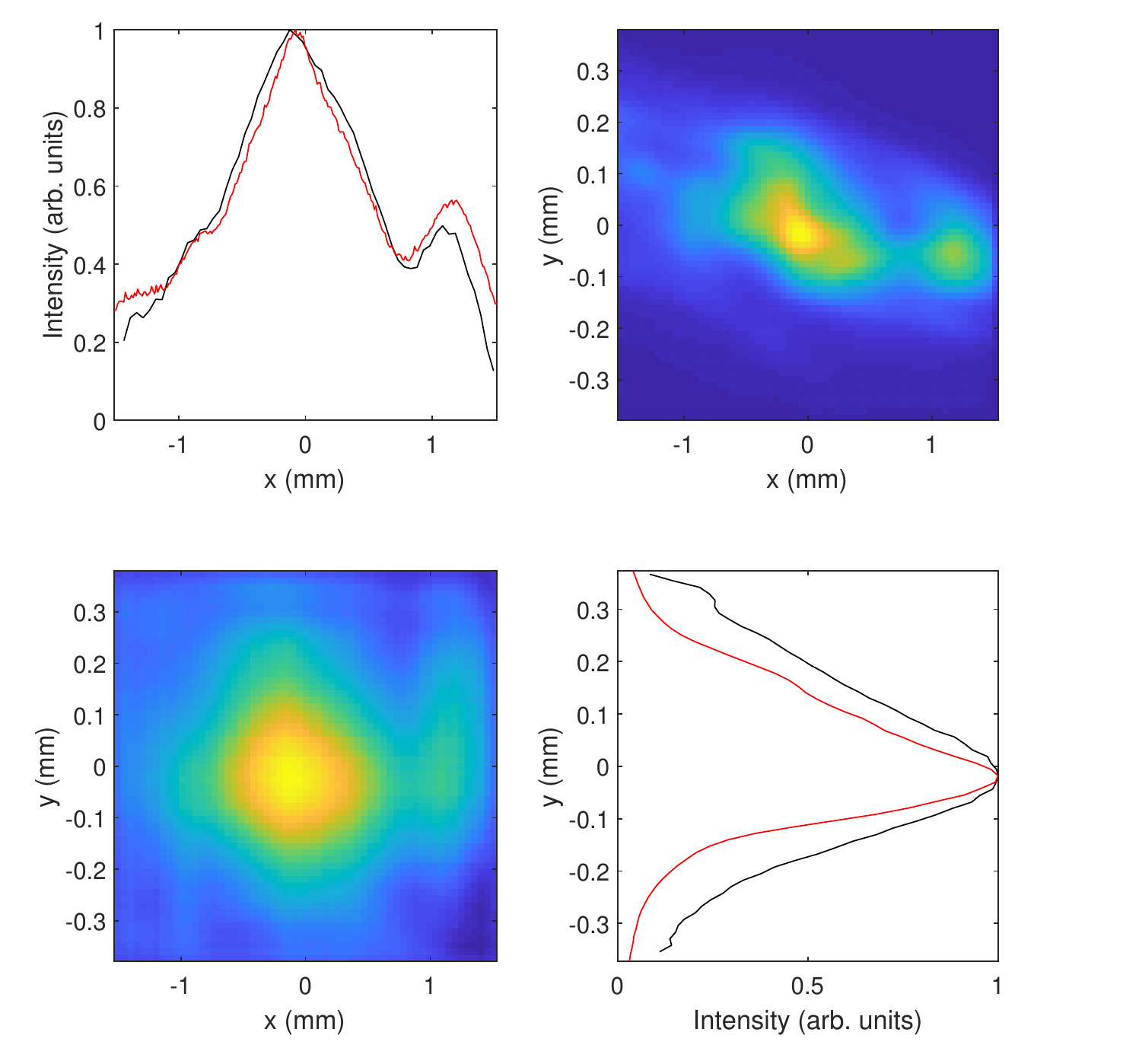} \\
(c) Main solenoid -150\,A, bucking coil -1.0\,A, &
(d) Main solenoid -150\,A, bucking coil -5.0\,A, \\
bunch charge 10\,pC &
bunch charge 10\,pC \\
 &
\end{tabular}
\caption{Comparison between beam images observed directly at SCR-02 and the reconstruction of the images from
phase space tomography in two degrees of freedom for a range of solenoid and bucking coil currents in the electron source
and different bunch charges in CLARA FE.  Within each subfigure, the plot at the top right shows the beam
image observed directly on SCR-02; the plot at bottom left shows the charge distribution in co-ordinate space
reconstructed from phase space tomography (a projection of the four-dimensional phase space onto the co-ordinate
plane) at the same location.  The plots at top left and bottom right in each subfigure show the charge density projected
onto the horizontal and vertical axes, respectively: black lines are from the tomography analysis and red lines are from the
directly observed image.
\label{figtomography4dimage}}
\end{center}
\end{figure*}

We can further validate the results by reconstructing the two-dimensional distribution in co-ordinate space
at the Reconstruction Point (by projecting the four-dimensional phase space distribution onto the co-ordinate
axes), and comparing this with the image that is observed directly.  Some examples for such comparisons are shown in
Fig.~\ref{figtomography4dimage}.  In general, we find
that the images reconstructed from phase space tomography reproduce reasonably well the general
shape and some of the more detailed features of the images that are observed directly.  However, the
tomography does not reveal the same level of detail as can be seen in the observed image.  This may
be due in part to the limited resolution of the tomography analysis: for the analysis presented here, we
used a phase space resolution of 69 points on each of the four axes (which was at the upper limit set by
the available computer memory).  However, it is also likely that measurement errors also play a role.
We note that the residuals of the fits to the images at the Observation Point are  typically very small (so
that there is no discernible difference between the directly-observed images at this point and the images
reconstructed from phase space tomography: see the example in Fig.~\ref{figtomography4dresiduals}).
However, there are systematic differences between the reconstructed images at SCR-02 (the
Reconstruction Point), and the images observed directly on that screen. In particular, the vertical size of 
the reconstructed beam (projecting the phase space distribution onto the vertical axis) is generally of
order 10\% larger than the vertical size of the image observed directly.  Work is in progress to
understand and correct the systematic errors: possible sources include calibration errors in the
quadrupoles and in the diagnostics used for collecting beam images.  It is important to have accurate
values for the quadrupole strengths, since the tomographic analysis depends on knowing the betatron
phase advance between the Reconstruction Point and the Observation Point, as well as the optical
functions at the Observation Point for given values of these functions at the Reconstruction Point.
Similarly, the analysis depends on accurate knowledge of the calibration factors of the diagnostic screens.
Hysteresis in the quadrupole magnets used to change the optics between the Reconstruction Point and
the Observation Point may also lead to errors in the analysis: to try to minimise hysteresis effects, the
quadrupoles were routinely degaussed (cycled) between scans, but the time taken for this procedure
made it impractical to degauss the quadrupoles at each point in a single scan.


\section{\label{sec:applications}Emittance and optics measurements under various machine conditions}



\subsection{Nominal machine settings}

Table~\ref{tablebeamparametercomparison} shows the emittance and optics parameters obtained under nominal
machine settings using the three different techniques discussed in the previous sections: three-screen measurements,
quadrupole scans, and phase space tomography. With the nominal machine settings, the electron source and linac
operate with the beam on-crest of the rf (i.e.~to give maximum beam acceleration for a given rf amplitude), with
amplitudes producing beam momentum 5\,MeV/c and 30\,MeV/c respectively.  The current in the bucking coil is set to
cancel the solenoid field on the  photocathode, and the laser intensity is set to give a bunch charge of 10\,pC.
The results in Table~\ref{tablebeamparametercomparison} are based on the same data set
(i.e.~the same set of beam images) in each case; the only difference between the different methods is in the way
that the data are analysed.  In principle, therefore, we expect to see good agreement between the values obtained
using different techniques.  The quadrupole scan and tomography techniques do indeed produce values in broad
agreement.  In the case of the three-screen analysis, however, the values found are significantly different from those
obtained using the other techniques. As discussed in Section~\ref{sec:threescreen}, this is probably because of the
complicated structure of the beam in phase space.  Under such conditions, the quadrupole scan and tomography
methods produce more reliable (and meaningful) results.

\begin{table*}
\begin{center}
\caption{Comparison of values for normalised emittances and Courant--Snyder parameters determined from different
analysis techniques. The results from analysis in four-dimensional phase space are for the normal mode emittances and
coupled optical functions (see Section~\ref{sec:emittancecalculation}).
The values for the three-screen analysis technique show the mean and standard deviation of the results of the analysis over
each point in the quadrupole scan, omitting points that do not return a real value for the emittance.  The three-screen analysis
method neglects any detailed structure in the beam distribution in phase space, and in this case leads to unreliable results.
\label{tablebeamparametercomparison}}
\begin{tabular}{| l | c | c | c || l | c | c | c |}
\hline
 \multicolumn{4}{| c ||}{Two-dimensional phase space} & \multicolumn{4}{ c |}{Four-dimensional phase space} \\
 \hline
 & \multirow{2}{*}{Three-screen} & \multirow{2}{*}{Quad scan} & \multirow{2}{*}{Tomography} & & \multirow{2}{*}{Quad scan} & Tomography & Tomography \\
 & & & & & & measurement & simulation \\
\hline
$\epsilon_{\mathrm{N},x}$ ($\mu$m) & 39.7$\pm$6.7 & 11.3 & 5.31 & $\epsilon_\mathrm{N,I}$ ($\mu$m) & 7.49 & 4.96  & 4.78 \\
$\beta_x$ (m) & 1.92$\pm$0.62 & 6.57  & 16.4 & $\beta^\mathrm{I}_{11}$ (m)& 8.84 & 19.8 & 20.8 \\
$\alpha_x$ & -0.231$\pm$0.060 & -1.37  & -1.64 & $-\beta^\mathrm{I}_{12}$& -1.51 & -2.03 & -2.30 \\
\hline
$\epsilon_{\mathrm{N},y}$ ($\mu$m) & 4.86$\pm$0.11 & 4.80 & 4.20 & $\epsilon_\mathrm{N,II}$ ($\mu$m) & 3.47 & 2.61 & 2.52 \\
$\beta_y$ (m) & 1.25$\pm$0.03 & 1.26 & 1.61 & $\beta^\mathrm{II}_{33}$ (m) & 1.79 & 1.29 & 1.30 \\
$\alpha_y$ & -1.53$\pm$0.21 & -1.80 & -1.75 & $-\beta^\mathrm{II}_{34}$ & -1.82 & -1.29 & -1.29 \\
\hline
\end{tabular}
\end{center}
\end{table*}

For the quadrupole scan and tomography analysis in four-dimensional phase space, the emittance and optics values given
are those for the normal mode quantities as described in Section~\ref{sec:emittancecalculation}.  The close agreement
with the uncoupled values (two-dimensional phase space) indicates that coupling is small in this case
(as expected from the machine settings).  The emittance and optics values in the case of the tomography analysis
are determined from the covariance matrix with elements calculated by averaging over the phase space density.

\subsection{Effect of varying bucking coil strength}

The electron source in CLARA FE is constructed so that the field from the solenoid can
be cancelled at the cathode by the field from a bucking coil.  If the current in the bucking coil is changed
from the value needed to achieve cancellation, electrons are emitted from the surface of the cathode in a non-zero
solenoid field: the effect is to introduce some coupling into the beam (as a result of non-compensated
azimuthal momentum), which can appear as changes in the beam emittances.  In particular, the individual normal
mode emittances will vary, though their product should remain constant as a function of the solenoid
field strength on the cathode \cite{derbenev1998,brinkmann2001}.
The difference between the normal mode emittances is expected to be minimised when
there is zero solenoid field on the cathode: with increasing field strength (parallel to the longitudinal axis,
in either direction) one emittance will increase while the other will decrease.  Tuning the machine
for optimum performance generally involves minimising the coupling, to achieve the smallest
possible emittance ratio \cite{zheng2019}, and characterising and understanding the coupling as a function of the
strength of the bucking coil is thus an important step in machine commissioning. Four-dimensional
phase space tomography offers a powerful tool for providing insight into coupling in the machine,
and was used to study the dependence of the phase space distribution on the strength of the
bucking coil.

\begin{figure}[th]
\begin{center}
\includegraphics[width=0.99\columnwidth,trim={5.5cm 12.5cm 4cm 6cm},clip]{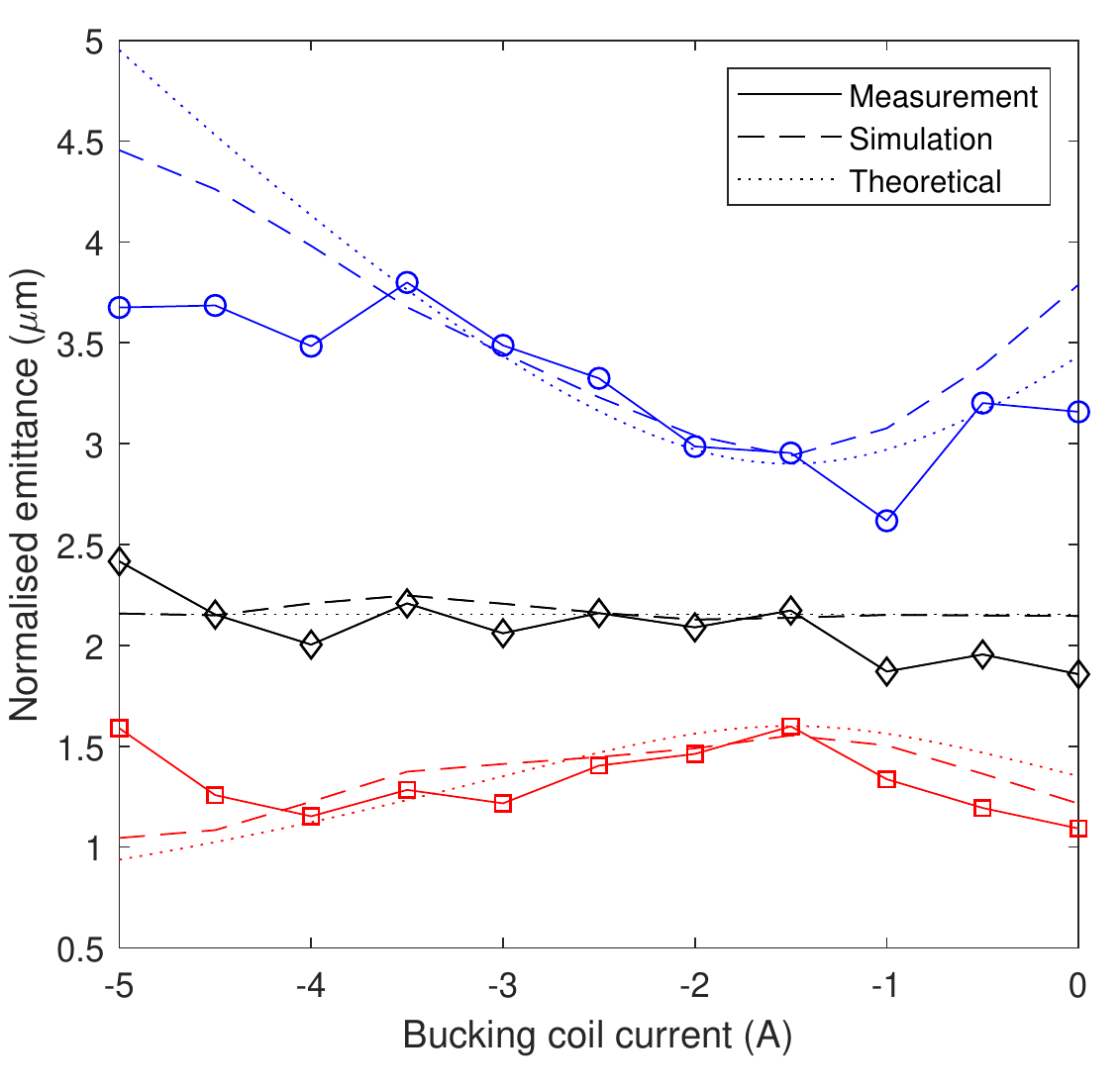}
\caption{Normalised normal mode emittances as a function of current in the bucking coil at the electron source,
with fixed bunch charge and main solenoid current.  The solid lines show experimental (measured)
values determined from four-dimensional phase space tomography (as described in Section~\ref{sec:tomography}),
the dashed lines show values from GPT simulations, and the dotted lines show values from a simple
theoretical model.  The upper and lower sets of lines (blue and red, respectively) show the normalised
normal mode emittances; the middle set of lines (black) show the geometric mean of the emittances.
Parameters in the simulations and theoretical model are chosen to fit the experimental results.
\label{beamemittancebcoilscan}}
\end{center}
\end{figure}

The normal mode emittances as a function of current in the bucking coil, found from four-dimensional
phase space tomography (as described in Section~\ref{sec:quadscan}) are shown in
Fig.~\ref{beamemittancebcoilscan}.  Although there is some variation in the product of the emittances
with changes in the current in the bucking coil, over a wide range the variation is small.  There is also some
indication of the expected behaviour of the individual emittances.  The difference between the emittances
is minimised for a bucking coil current of approximately -1.5\,A: this is somewhat different from the nominal
value of -2.2\,A for cancelling the field on the cathode.  The reason for the discrepancy is being investigated.
Note that before collecting data over the range of bucking coil currents,
the bucking coil was degaussed with the intention of improving the agreement between the
cathode field calculated from a computer model of the electron source and the field that was
actually produced for a given current.
It is also worth noting that the time taken for data collection over the full range of bucking
coil currents took several hours, and it is likely that some variation in machine parameters
(such as rf phase and amplitude in the electron source and linac) occurred over this time.

Also shown in Fig.~\ref{beamemittancebcoilscan} are results from a GPT simulation and from a simple theoretical
model: these are included in the figure to illustrate the expected behaviour of the normal mode emittances as a
function of the solenoid field on the cathode, and are not intended to show results from an accurate machine model.
For the simulations, we use parameters for the electron source corresponding to those
in CLARA FE, but with the field from the bucking coil scaled to cancel the solenoid field on the cathode
for a current of -1.5\,A in the the bucking coil (rather than the nominal -2.2\,A).  Also, the initial distribution of
particles in phase space is chosen to give emittances (with zero solenoid field on the cathode) corresponding to
the experimental measurements.  This requires values for the beam size and divergence at the cathode that
are significantly different from the values believed to be appropriate for CLARA FE, by a factor of two in spot size,
and up to an order of magnitude in divergence; however, it should be remembered that in the simulation, the emittances
are calculated immediately after the electron source, whereas the measurements are made in a section of beamline
downstream of the linac and numerous other components.  Effects (that are not yet well characterised) between the
electron source and the measurement section are likely to lead to some increase in emittance. The GPT
and theoretical results are therefore included in Fig.~\ref{beamemittancebcoilscan} purely to illustrate
the expected behaviour of the emittances as functions of the strength of the solenoid field on the cathode,
rather than as a direct comparison of an accurate computational model with the experimental results.

Also shown in Fig.~\ref{beamemittancebcoilscan} are results from a simplified theoretical
(analytical) model.  This is based on an assumed beam phase space distribution
at the cathode, i.e.~immediately after photoemission.  If there is no magnetic field on the cathode, then
the covariance matrix is characterised by an emittance and beta function
in each transverse direction:
\begin{equation}
\Sigma = \left( \begin{array}{cccc}
\beta_x \epsilon_x & 0 & 0 & 0 \\
0 & \frac{\epsilon_x}{\beta_x} & 0 & 0 \\
0 & 0 & \beta_y \epsilon_y & 0 \\
0 & 0 & 0 & \frac{\epsilon_y}{\beta_y}
\end{array}\right).
\end{equation}
A solenoid field of strength $B_0$ on the cathode can be represented by a vector
potential:
\begin{equation}
\mathbf{A} = \left( -\frac{1}{2}B_0 y, \frac{1}{2} B_0 x, 0 \right),
\end{equation}
so that the canonical conjugate momenta $p_x$ and $p_y$ become:
\begin{eqnarray}
p_x & = & \frac{\gamma m v_x + \frac{1}{2}e B_0 y}{P_0}, \\
p_y & = & \frac{\gamma m v_y - \frac{1}{2}e B_0 x}{P_0},
\end{eqnarray}
where $m$ and $e$ are the mass and magnitude of the charge of the electron, $v_x$
and $v_y$ are the transverse horizontal and vertical components of the velocity, and
$P_0 = \beta_0\gamma_0 mc$ is the reference momentum (which can be chosen arbitrarily).
The covariance matrix then becomes:
\begin{equation}
\Sigma = \left( \begin{array}{cccc}
\beta_x \epsilon_x & 0 & 0 & \eta \beta_x \epsilon_x \\
0 & \frac{\epsilon_x}{\beta_x} + \eta^2 \beta_y \epsilon_y & -\eta \beta_y \epsilon_y & 0 \\
0 & -\eta \beta_y \epsilon_y & \beta_y \epsilon_y & 0 \\
\eta \beta_x \epsilon_x & 0 & 0 & \frac{\epsilon_y}{\beta_y} + \eta^2 \beta_x \epsilon_x
\end{array}\right),
\label{flatbeamcovariancematrix}
\end{equation}
where:
\begin{equation}
\eta = \frac{e B_0}{2 P_0}.
\end{equation}
Finally, from the covariance matrix (\ref{flatbeamcovariancematrix}), we find (using the methods described in
Section~\ref{sec:emittancecalculation}) that the normal mode emittances are given by:
\begin{equation}
\epsilon_{\mathrm{I},\mathrm{II}} = \sqrt{\chi \pm \sqrt{\chi^2 - \epsilon_x^2 \epsilon_y^2}},
\end{equation}
where:
\begin{equation}
\chi = \frac{\epsilon_x^2 + \epsilon_y^2}{2} +2 \eta^2 \beta_x \epsilon_x \beta_y \epsilon_y.
\end{equation}
The normalised emittances ($\epsilon_\mathrm{N,I} = \beta_0 \gamma_0 \epsilon_\mathrm{I}$, and
similarly for $\epsilon_\mathrm{N,II}$) remain constant during acceleration of particles in the rf field of
the electron source (and in the linac).  To apply this model to CLARA FE, giving the results shown in
Fig.~\ref{beamemittancebcoilscan}, the initial beam size and divergence are chosen
to fit the emittances at their closest approach: the values used are close to those used in the GPT
simulation.  We also assume that $\eta = 0$ for a bucking coil current of -1.5\,A, and scale the
dependence of $\eta$ on the field in the bucking coil so as to match the experimental curves.
However, we again emphasise that the results from the theoretical model and the GPT simulation
are included only to give an illustration of the expected behaviour, and are not directly comparable
with the experimental results.

\begin{figure*}[th]
\begin{center}
\includegraphics[width=0.495\columnwidth,trim={3.3cm 4.1cm 11cm 18cm},clip]{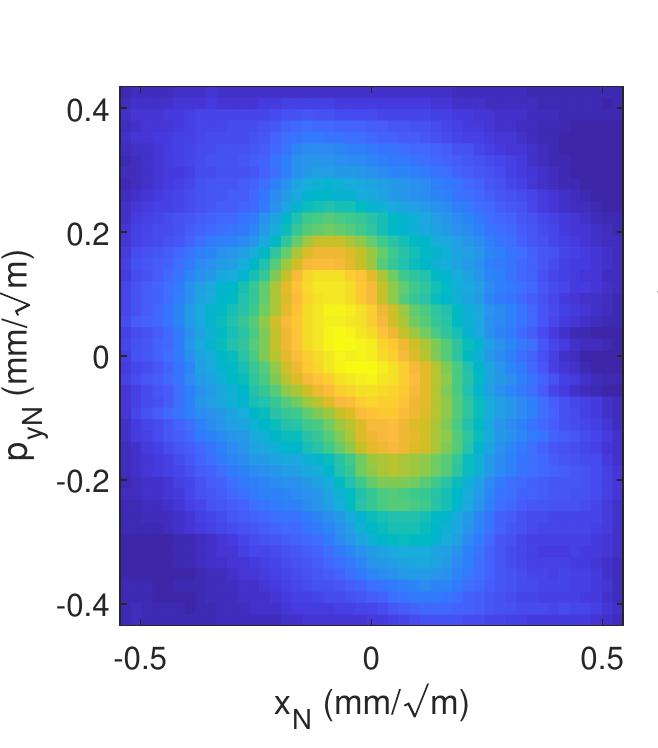}
\includegraphics[width=0.495\columnwidth,trim={3.3cm 4.1cm 11cm 18cm},clip]{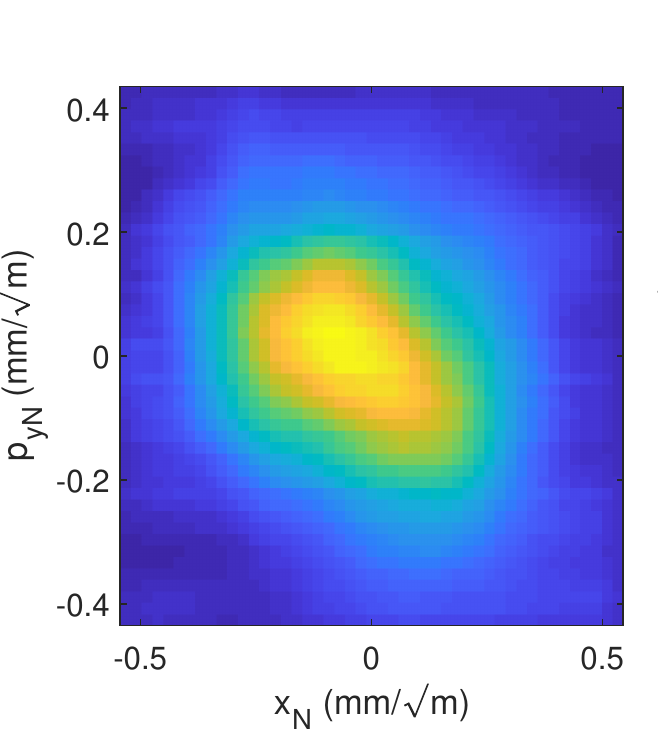}
\includegraphics[width=0.495\columnwidth,trim={3.3cm 4.1cm 11cm 18cm},clip]{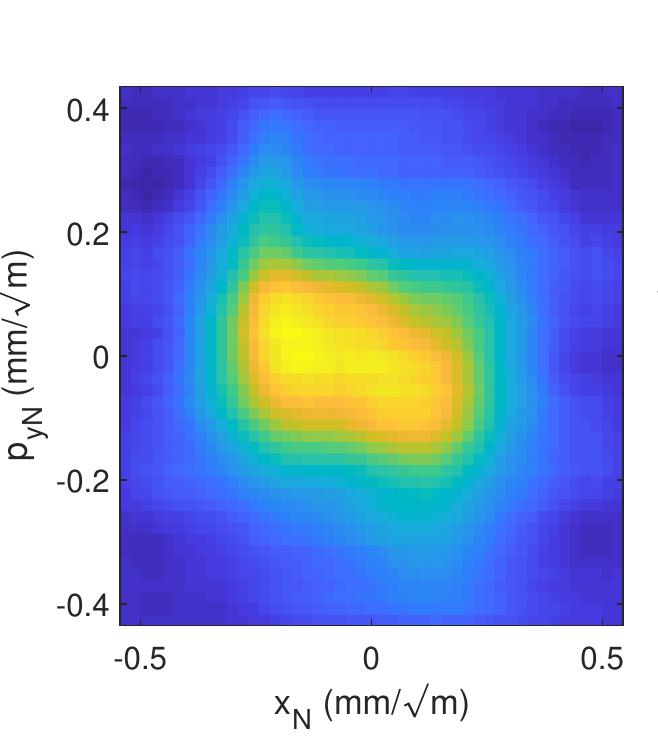}
\includegraphics[width=0.495\columnwidth,trim={3.3cm 4.1cm 11cm 18cm},clip]{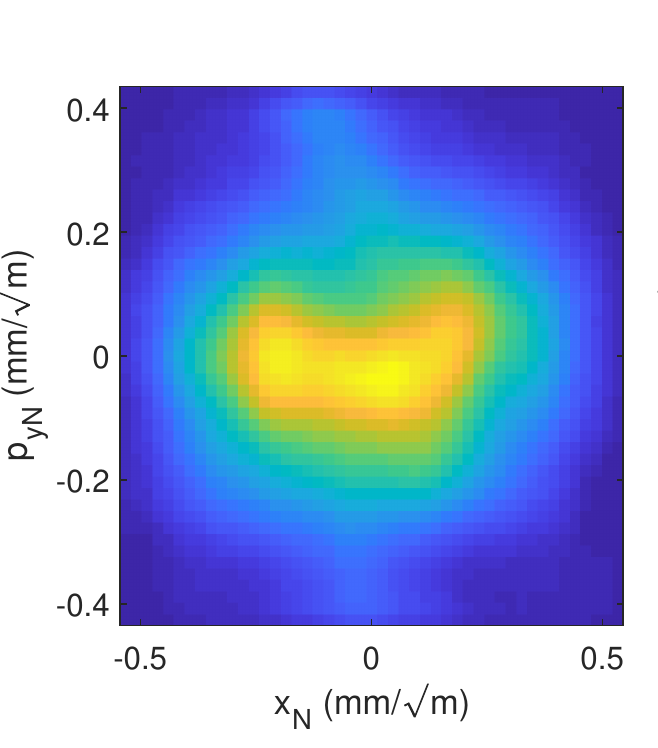}
\includegraphics[width=0.495\columnwidth,trim={3.3cm 4.1cm 11cm 18cm},clip]{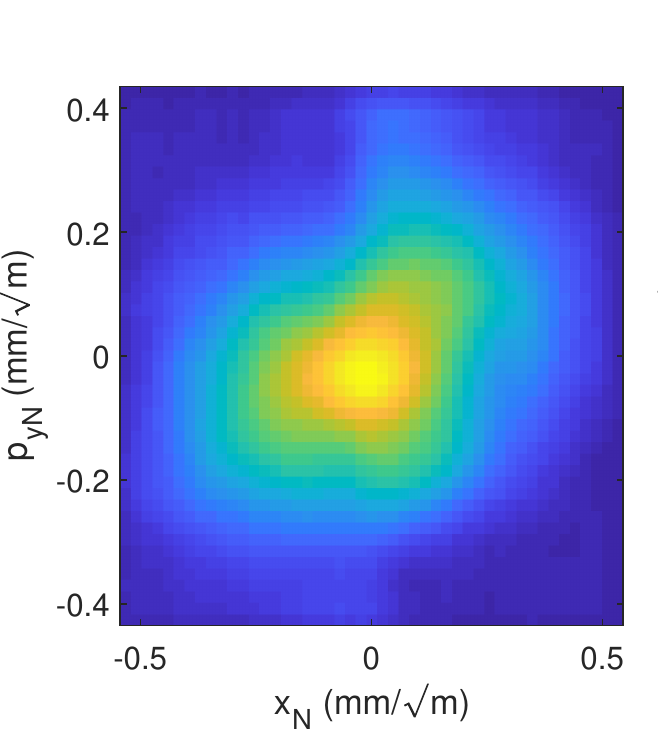}
\includegraphics[width=0.495\columnwidth,trim={3.3cm 4.1cm 11cm 18cm},clip]{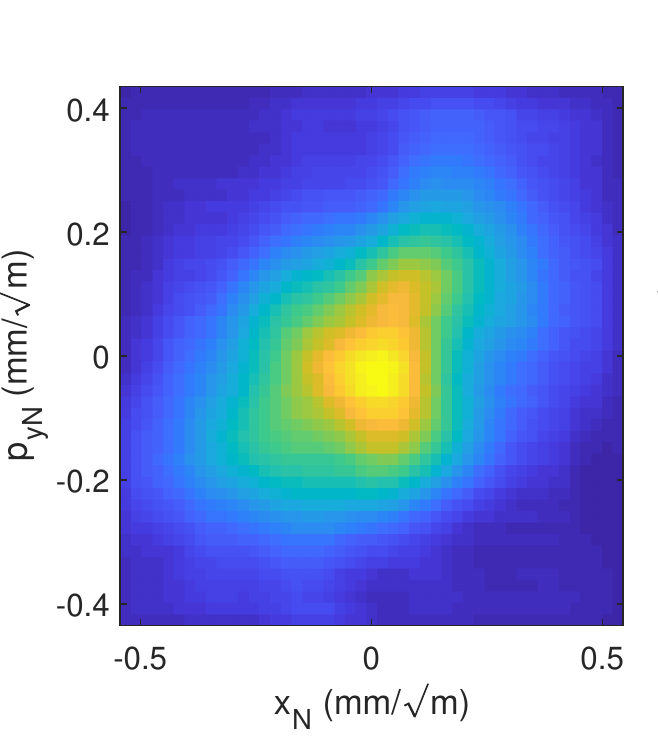}
\includegraphics[width=0.495\columnwidth,trim={3.3cm 4.1cm 11cm 18cm},clip]{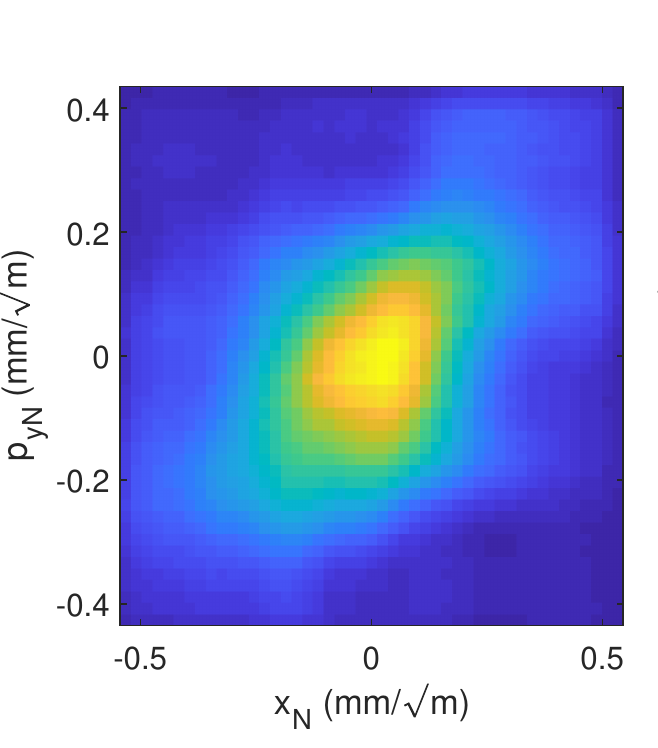}
\includegraphics[width=0.495\columnwidth,trim={3.3cm 4.1cm 11cm 18cm},clip]{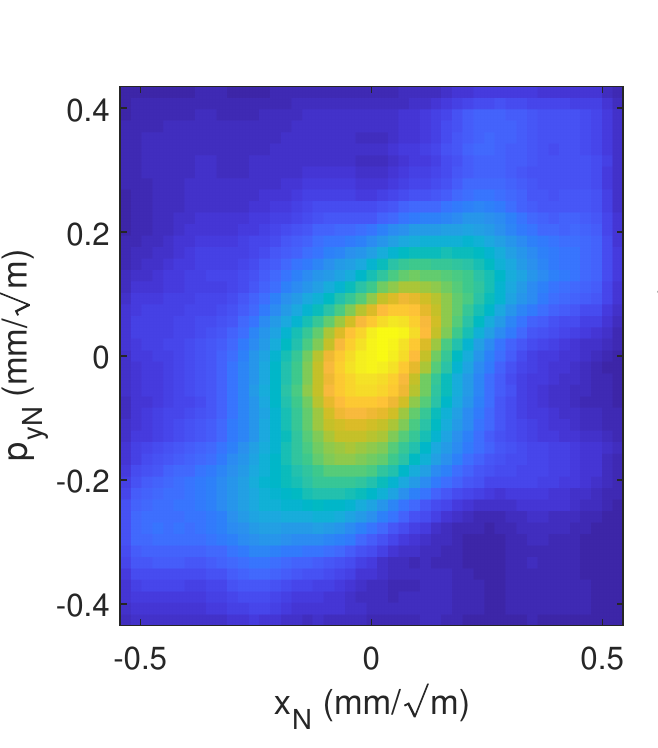}
\caption{Projection of the beam distribution in normalised phase space onto the $x_\mathrm{N}$--$p_{y\mathrm{N}}$ plane, reconstructed from
four-dimensional phase space tomography, showing variation of coupling with bucking coil current.  From left to right, top to bottom,
the bucking coil current increases from 0\,A (top left) to -3.5\,A (bottom right) in steps of -0.5\,A.  The correlation
(tilt) between horizontal co-ordinate $x_\mathrm{N}$ and the vertical momentum $p_{y\mathrm{N}}$ indicates the strength of the coupling.
\label{beamcouplingpxpy}}
\end{center}
\end{figure*}

Direct inspection of the phase space distribution provides a further indication of how the coupling changes
with the current in the bucking coil.  For example, Fig.~\ref{beamcouplingpxpy} shows the projection onto
the $x$--$p_y$ plane of the four-dimensional phase space (reconstructed from the tomography measurements)
for different values of the current in the bucking coil.  The ``tilt'' on the distribution corresponds to a correlation
between the horizontal co-ordinate and vertical momentum, and indicates the coupling: we see that this changes sign as
the bucking coil current is varied from 0\,A to -3.5\,A.  The tilt (and hence the coupling) vanishes for a current of
around $-1.5$\,A, which is consistent with the current required to minimise the difference between the normal mode
emittances.

\begin{figure*}[th]
\begin{center}
\includegraphics[width=0.80\textwidth,trim={0.8cm 8.5cm 1cm 8cm},clip]{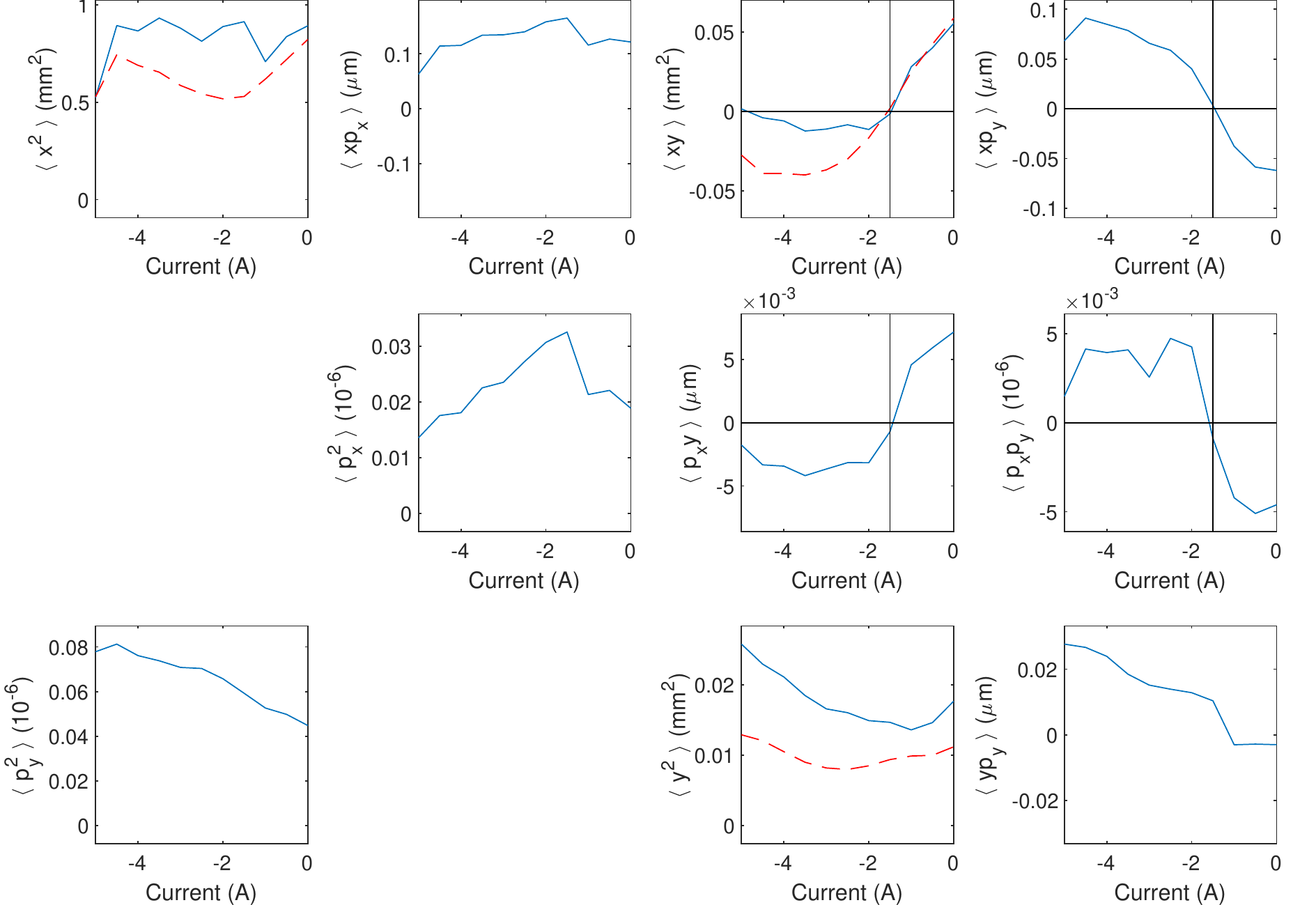}
\caption{Elements of the covariance matrix at SCR-02 (the Reconstruction Point) as functions of current in the bucking coil.
Blue solid lines show the values determined from four-dimensional phase space tomography; red dashed lines show the values
calculated from the directly observed image of the beam on SCR-02. Note that in the case of the tomography analysis, the
elements describing coupling between the transverse degrees of freedom (shown in the top right $2\times 2$ block diagonal)
vanish for a bucking coil current of approximately -1.5\,A: this is consistent with the current at which the difference between
the normal mode emittances is minimised (as shown in Fig.~\ref{beamemittancebcoilscan}).
\label{beamcovmatrixbcoilscan}}
\end{center}
\end{figure*}

A more complete characterisation of the coupling is given in Fig.~\ref{beamcovmatrixbcoilscan}, which shows the
elements of the covariance matrix at SCR-02 (the Reconstruction Point) as functions of current in the bucking coil.
Coupling between motion in the horizontal and vertical directions is indicated by non-zero values of the elements
in the $2\times 2$ top-right block diagonal.  All these elements vanish for bucking coil currents of around -1.5\,A.
We do not expect this to correspond exactly to the bucking coil current that minimises the separation between
the normal mode emittances, since after leaving the cathode (in zero longitudinal magnetic field) the particles
then pass through a section of main solenoid field, not cancelled by the bucking coil.  The main solenoid field
introduces some coupling in the beam, characterised by non-zero elements off the $2\times 2$ block diagonals
in the covariance matrix.  However, tracking simulations in GPT suggest that in the case of CLARA FE, the coupling
in the covariance matrix introduced by the part of the main solenoid not cancelled by the bucking coil is
small: the coupling in the covariance matrix is minimised at a current within about 0.2\,A of the current that
gives the closest approach of the normal mode emittances (see Fig.~\ref{beamemittancebcoilscan}).

%
%

\subsection{Effects of varying main solenoid strength and bunch charge}

Although space-charge effects in CLARA FE are negligible in the section of beamline where
the emittance and optics measurements are made (with beam momentum around 30\,MeV/c),
space-charge forces can play a significant role in the electron source, depending on the
bunch length and the total bunch charge.
In the studies reported here, the photocathode laser was operated with pulse length of 2\,ps:
in this regime, space-charge effects are expected to be weak, even at the higher bunch charges
(up to 250\,pC) achievable in the machine.  However, screen images suggested
some significant variation in beam parameters even at lower bunch charges.
It is planned in the future to use phase space tomography for rigorous studies of the impact of
bunch charge (and other parameters) on beam properties; but so far, the limited
time available for collecting quadrupole scan data, together with some variability in the machine conditions,
has made it impractical to make detailed, systematic measurements.   Nevertheless, to provide some
information on beam behaviour, quadrupole scans were performed for bunch charges of 10\,pC,
20\,pC and 50\,pC, and for a reduced main solenoid current of 125\,A, as well as the nominal 150\,A.
Some of the results from analysis of these quadrupole scans using phase space tomography in two
degrees of freedom are shown in Fig.~\ref{figtomography4dimage}, which compares the
reconstructed beam image at SCR-02 with the image observed directly on this screen.

The images in Fig.~\ref{figtomography4dimage} also indicate significant detailed structure
in the beam distribution, depending on main solenoid current, bucking coil current and bunch charge.
This is apparent both from the image observed directly at the Reconstruction Point, and from the
phase space distribution constructed from four-dimensional tomography.  In such cases, the
phase space cannot accurately be characterised simply by the emittances and optical functions that
describe the covariance matrix.  Nevertheless, to allow some comparison, we calculate the
normal mode emittances and optical functions, using the method described in
Section~\ref{sec:emittancecalculation}: the values of the normal mode emittances and selected optical
functions are shown in Table~\ref{tablechargescan}.  Although there are indications of some
patterns (for example, an increase in emittance with bunch charge) no firm conclusions can be
drawn because machine conditions between different quadrupole scans were not accurately
reproducible. Nevertheless, the measurements that have been made demonstrate the potential
value of four-dimensional phase space tomography for developing an understanding of the beam
physics in a machine such as CLARA FE, and for tuning the machine for optimum performance.

\begin{table*}
\begin{center}
\caption{Beam emittances and optics parameters with different bunch charges and main solenoid strengths,
with bucking coil current at the nominal -2.2\,A.  Note that measurements for some settings of the bunch
charge and main solenoid strength were repeated (in particular: 10\,pC bunch charge and -125\,A main
solenoid current; and 50\,pC bunch charge and -150\,A main solenoid current), to indicate the
reproducibility of the tomography analysis for given machine settings.
\label{tablechargescan}}
\begin{tabular}{| c | c | c | c | c | c | c | c |}
\hline
Bunch charge & Main solenoid current & $\epsilon_\mathrm{N,I}$ & $\epsilon_\mathrm{N,II}$ & $\beta^\textrm{I}_{11}$ & $\beta^\textrm{II}_{33}$ & $\beta^\textrm{I}_{33}$ & $\beta^\textrm{II}_{11}$ \\
(pC) & (A) & ($\mu$m) & ($\mu$m) & (m) & (m) & (m) & (m) \\
\hline
10 & -125 & 6.02 & 3.09  & 12.5 & 1.92 & 0.394 & 2.20 \\
10 & -125 & 10.2 & 6.33 & 11.8 & 1.45 & 0.368 & 3.39 \\
10 & -150 & 3.75 & 1.37 & 8.92 & 0.417 & 0.157 & 8.37 \\
20 & -125 & 13.7 & 5.93 & 15.8 & 3.52 & 0.287 & 3.97 \\
20 & -150 & 8.88 & 3.33 & 14.3 & 1.10 & 0.065 & 1.50 \\
50 & -150 & 8.98 & 4.47 & 18.8 & 0.981 & 0.121 & 3.28 \\
50 & -150 & 9.25 & 4.59 & 18.8 & 1.00 & 0.114 & 2.88 \\
\hline
\end{tabular}
\end{center}
\end{table*}

%
%


\section{\label{sec:conclusions}Summary and conclusions}



We have presented the first experimental results from four-dimensional phase space
tomography in an accelerator.  The beam emittance and optical properties obtained from
phase space tomography have been compared with results obtained using more commonly
employed techniques, such as three-screen analysis and quadrupole scans.  The comparisons
show that where there are detailed structures in the beam distribution in phase space (so
that the distribution cannot be described, for example, by a simple Gaussian), three-screen
and quadrupole scan analysis provide limited, and not always reliable, information.  The
difficulty in the case of a non-Gaussian beam distribution, is that a detailed description of the
distribution cannot simply be given in terms of a small number of parameters (emittance and
Courant--Snyder parameters).  Phase space tomography overcomes this limitation by
providing the beam density at a number of points in phase space.

Our results for the phase space tomography analysis and the comparisons with other
methods are supported by simulation studies.  The results of the tomography analysis have been
validated by comparing (for example) the beam image at the entrance of the
measurement section of the beamline (the Reconstruction Point) obtained from a projection of
the measured four-dimensional phase space, with the beam image observed directly on a screen at this
point.  In general, the agreement suggests that four-dimensional phase space tomography
is providing a useful representation of the beam properties, though the image reconstructed
from tomography lacks the same resolution as the image observed directly.  There is also
evidence for systematic errors in the measurement that need to be properly understood.

A benefit of four-dimensional phase space tomography (compared to tomography in two-dimensional
phase space) is that the technique provides detailed information on coupling in the beam.  This can be
important for tuning a machine such as CLARA FE, for example, where solenoids are used
to provide focusing for the beam, but it is desirable to minimise the coupling that can be
introduced by those solenoids.  Information on coupling can be obtained by applying the quadrupole
scan method in two (transverse) degrees of freedom; but information obtained in this way may not
be accurate or reliable if there is detailed structure in the beam distribution.

The main drawback of the tomography analysis is that collection of the data may be
a time-consuming procedure.  In cases where the beam distribution in phase space is smooth
and without significant detailed structure (so that it can be well characterised by the
emittance and optical functions) then the three-screen or quadrupole scan techniques,
using a limited set of observations, may provide sufficient information 
for machine tuning and optimisation relatively quickly.  Phase space tomography generally
requires data from a larger number of observations, but depending on the level
of detail or accuracy required, it may be possible to minimise the number of points in the
quadrupole scan used to provide the data: the limits of the
technique have still to be rigorously explored, and will likely depend on the specific machine
to which it is applied.

Regarding practical application of phase space tomography, it is worth mentioning that
the requirements in terms of beamline design and diagnostics capability are not demanding.
In CLARA FE, the diagnostics section consists of a short (1.661\,m) section of beamline between
two transverse beam profile monitors, and containing three (adjustable strength) quadrupoles.  The
design of this section was developed before detailed plans were prepared for phase space tomography
studies, and there is limited flexibility in optimising the phase advances and optical functions
over the length of the diagnostics section.  Nevertheless, it was possible to identify sets of
quadrupole strengths to provide the observations necessary for the analysis and results
presented here.

So far, we have used an algebraic reconstruction technique for the phase space tomography.
This technique has the advantage (compared to other tomography algorithms) of ease of implementation and
flexibility in terms of the input data.  However, it is possible that different algorithms may provide better
(more accurate, or more detailed) results, and we hope to explore the possible benefits
and limitations of alternative tomography methods.  A particular issue with tomography
in four-dimensional phase space is the demand on computer memory for processing the
data and storing the results, especially at high resolution in phase space.  However, because
of the nature of the problem, the memory requirements will almost inevitably scale with
the fourth power of the phase space resolution, and it seems unlikely that other tomography
methods would provide significant benefit in this respect.  It is possible that more sophisticated
computational techniques may allow some reduction in the memory requirements for a given
resolution, e.g.~\cite{cosme2018}.

While improvements and refinements in the technique are planned, the results so far show
that four-dimensional phase space tomography is a useful technique for detailed beam
characterisation and for machine tuning and optimisation.  It is hoped that further studies
will include investigation of space-charge effects in the electron source and beam dynamics effects
(such as wake fields) in the linac.

\acknowledgements

We would like to thank our colleagues in STFC/ASTeC at Daresbury Laboratory
for help and support with various aspects of the simulation and experimental
studies of CLARA FE.

This work was supported by the Science and Technology Facilities Council, UK,
through a grant to the Cockcroft Institute.


\end{document}